\documentclass[a4paper,10pt]{article}
\usepackage[round]{natbib}
\usepackage{subfig}
\usepackage[utf8x]{inputenc}
\usepackage{threeparttable}
\pdfoutput=1
\usepackage{verbatim}
\usepackage{epsfig}
\usepackage{epsf}
\usepackage{epstopdf}
\usepackage{graphicx}
\usepackage{textcomp}
 \usepackage{amsmath}
\usepackage{color}
\usepackage{multirow}
\definecolor{Blue}{rgb}{0.3,0.3,0.9}
\definecolor{red}{rgb}{1,0,0}
\usepackage{pgfplots}
\usepackage{tikz}
\usepackage{caption}

\newcommand{\vectwo}[2]
{
   \begin{pmatrix} #1 \\ #2 \end{pmatrix}
}

\title{\textbf{\fontsize{15}{15}\selectfont The Gabor-Einstein Wavelet: A Model for the Receptive Fields of V1 to MT Neurons}}\author{\smallskip \textbf{\fontsize{11}{11}\selectfont Stephen G. Odaibo$^{1,2}$}\\ \textbf{\fontsize{10}{10}\selectfont M.S.(Math), M.S.(Comp. Sci.), M.D.}\\ \newline\newline \small{\textsl{\fontsize{9}{9}\selectfont $^1$Quantum Lucid Research Laboratories,}}\\ \small{\textsl{\fontsize{9}{9}\selectfont Computational Visual Neuroscience Group}}\\ \small{\color{blue}{\fontsize{7}{7}\selectfont stephen.odaibo@qlucid.com}}\\\newline\newline \\\small{\textsl{\fontsize{9}{9}\selectfont $^2$Howard University Hospital, Ophthalmology}}\\  \small{\textsl{\fontsize{7}{7}\selectfont Washington, D.C., U.S.A.  }} }
\date{}

\date{}

\begin{document}

\maketitle

\begin{abstract}

Our visual system is astonishingly efficient at detecting moving objects. This process is mediated by the neurons which connect the primary visual cortex (V1) to the middle temporal (MT) area. Interestingly, since Kuffler's pioneering experiments on retinal ganglion cells, mathematical models have been vital for advancing our understanding of the receptive fields of visual neurons. However, existing models were not designed to describe the most salient attributes of the highly specialized neurons in the V1 to MT motion processing stream; and they have not been able to do so.  Here, we introduce the Gabor-Einstein wavelet, a new family of functions for representing the receptive fields of V1 to MT neurons. We show that the way space and time are mixed in the visual cortex is analogous to the way they are mixed in the special theory of relativity (STR). Hence we constrained the Gabor-Einstein model by requiring: (i) relativistic-invariance of the wave carrier, and (ii) the minimum possible number of parameters. From these two constraints, the sinc function emerged as a natural descriptor of the wave carrier. The particular distribution of lowpass to bandpass temporal frequency filtering properties of V1 to MT neurons (Foster et al 1985; DeAngelis et al 1993b; Hawken et al 1996) is clearly explained by the Gabor-Einstein basis. Furthermore, it does so in a manner innately representative of the motion-processing stream's neuronal hierarchy. Our analysis and computer simulations show that the distribution of temporal frequency filtering properties along the motion processing stream is a direct effect of the way the brain jointly encodes space and time. We uncovered this fundamental link by demonstrating that analogous mathematical structures underlie STR and the joint cortical encoding of space and time. This link will provide new physiological insights into how the brain represents visual information.


\end{abstract}
\hspace{-1.5mm}\indent\indent {{\fontsize{7}{7}\selectfont {\bf Keywords:} Visual cortex, V1, MT, Receptive field, Gabor, Response properties}}
\begin{center}
\line(1,0){300}
\end{center}

\newpage

\section{Introduction}

\subsection{Aim}

To introduce a model which innately and efficiently represents the salient properties of receptive fields of neurons in the V1 to MT motion processing stream. 

\subsubsection{Specific Aim}

As one traces the neural pathway connecting the input layers of V1 to area MT, neurons become increasingly specialized for motion detection. In ascending order, attributes such as orientation selectivity, direction selectivity, speed tuning, increasing preferred speeds, component spatial frequency selectivity, and pattern (plaid) spatial frequency selectivity are sequentially acquired. Furthermore, the temporal frequency filtering properties of the neuronal population changes in a particular way along this pathway. Specifically, the proportion of bandpass temporal frequency filtering neurons to lowpass temporal frequency filtering neurons increases. Existing receptive field models do not represent this fundamental emergent property. Our specific aim is to introduce and describe a model for the receptive fields of V1 to MT neurons which innately and efficiently represents the aforementioned properties.

\subsection{Motivation and Background}

The detection of moving objects in our environment is both critical to our experience of the world and to our very survival. Imagine a hunter in rapid pursuit of a small agile animal. Both predator and prey detect abrupt changes in velocity occurring over an interval of less than a tenth of a second. The precision of our visual motion detection system is undoubtedly striking.

The middle temporal (MT) area of the mammalian brain plays a key role in the visual processing of motion. The importance of area MT in motion processing is supported by a growing wealth of electrophysiological evidence which was initiated by Dubner and Zeki's 1971 report \citep*{duze1971,ze1974,ze1980,vama1981,mava1983,mava1983b,feka1984,unde1986,mane1987,roal1989,mone1996,smma2005}. For purposes of motion processing, the main input signals into MT originate in specialized cells of the primary visual cortex (V1). 

Kuffler's early studies of retinal ganglion cell response properties led to similar studies of simple cortical cells by Hubel and Wiesel. Together, their work generated great interest in neuronal receptive fields~\citep*{ku1953,hu1957,hu1959,huwi1959,ma1980,da1985}. The functional forms of these receptive fields are at once beautifully simple yet enormously complex. Hence mathematical models have been used in step with electrophysiological studies to advance our understanding of their properties. Initially, focus was predominantly on their spatial structure~\citep*{ma1980,ga1946}. Now, however, their temporal structure is increasingly studied in tandem. In particular, for most neurons in the V1 to MT processing stream, it is now appreciated that spatial and temporal features cannot be studied separately. They are spatiotemporally inseparable entities. In particular, motion is encoded by orientation in the spectral domain, and spatiotemporally-oriented filters are therefore motion detectors~\citep{waah1983,waah1985,he1987}. Therefore at first glance it may seem an easy matter to mathematically model motion detecting neurons. The challenge, however, is to develop physiologically sound receptive field models which reflect the hierarchical structure of the motion processing stream. Various models do exist which are spatiotemporally-oriented filters, and are therefore motion detectors from a mathematical standpoint~\citep*{adbe1985,qi1994,qian1994,qian1997,qifr2009}. However, they fail to represent one of the most salient characterizing attributes of the motion processing stream: the lowpass to bandpass distribution of temporal frequency filtering properties along the V1 to MT specialization hierarchy. This feature is discussed further in the next subsection.

 In addition to the above temporal frequency filter property profile, the neurons of the V1 to MT motion processing stream have certain characteristics which set them apart and prescribe bounds on models of their spatiotemporal receptive field structure. Namely, they are tuned to specific orientations, directions, and speeds \citep*{mone1996}; and they are more likely than other V1 cells to exhibit ``end-stopping'' phenomena~\citep*{scha2001}. Of the V1 cells, the subclass with direct synaptic projections from V1 to MT are the most specialized towards motion and are thought to be the main channels of motion substrates incident into MT neurons. For instance, Churchland et al showed that V1 and MT neurons lose direction selectivity for similar values of spatial disparity, suggesting V1 is MT's source of direction selectivity~\citep*{chpr2005}. Additionally, these V1 to MT neurons cluster motion substrate properties such as strong direction selectivity and tuning to high speeds. For example, Foster et al found that highly direction selective neurons in the macaque V1 and V2 were more likely to be tuned to higher temporal frequencies and lower spatial frequencies, i.e. higher speeds~\citep*{foga1985}. In other words, neurons that are highly direction selective are more likely to also be tuned to higher speeds. Such neurons are higher up in the V1 to MT hierarchy. Similarly, McLean and Palmer studied direction selectivity in cat striate cortex, and found that space-time inseparable cells were direction selective while space-time separable cells were not direction selective~\cite{mcpa1989}. Again reflecting the hierarchy of the motion processing stream.

The distribution of some distinct anatomical and histological features have been shown to correlate with the motion-specialization hierarchy. For instance, Shipp and Zeki's retrograde tracer studies showed that the majority of direct V1 to MT projecting neurons originate within layer 4B~\citep*{shze1989}. Based on their results, they suggested that layer 4B contained a functionally subspecialized anatomically segregated group whose members each have direct synaptic connections with MT cells. The exact properties of the V1 to MT projectors are likely to be intermediate between those of cells in V1 layer 4C$\alpha$ and MT neurons. This is because input into MT appears to be largely constituted of magnocellular predominant streams from 4C$\alpha$, while parvocellular predominant streams may play a much less role ~\citep*{lihu1988,mane1990,yasa2001}. Progressive specialization along the V1 to MT motion processing stream is also observable in the morphological characteristics of the neurons. For instance, distinguishing attributes of the V1-MT or V2-MT projectors such as size, arborization patterns, terminal bouton morphology, and distribution have been observed ~\citep*{ro1989,ro1995,anbi1998,anma2002}. Sincich and Horton demonstrated that layer 4B neurons projecting to area MT were generally larger than those projecting to layer V2. Overall, neurons in the V1 to MT motion processing stream are highly and progressively specialized. Hence description of their receptive fields requires adequately specialized models which reflect not only their individual attributes, but also the emergent hierarchical properties of the network. Next we discuss one of the most fundamental of such emergent network properties: the particular distribution of temporal frequency filtering types along the stream~\citep{foga1985,deoh1993b,hash1996}.

\subsection{Temporal Frequency Filtering Property Distribution}

Hawken et al found that direction-selective cells were mostly bandpass temporal frequency filters, while cells which were not direction-selective were equally distributed into bandpass and lowpass temporal frequency filtering types \citep*{hash1996}. Foster et al found a similar phenomenon in macaque V1 and V2 neurons. V1 neurons were more likely to be lowpass temporal frequency filters, while their more specialized downstream heirs, V2 neurons, were more likely to be bandpass temporal frequency filters~\citep*{foga1985}. Less specialized neurons located anatomically upstream (lower down in the hierarchy) are more likely to have lowpass temporal frequency filter characteristics, while more specialized neurons located anatomically downstream (higher up in the hierarchy) are more likely to display bandpass temporal frequency filter characteristics. For example, LGN cells are at best only weakly tuned to direction and orientation (~\citet{xuic2001},~\citet*{femi2000}) and are hence equally distributed into lowpass and bandpass categories. Layer 4B V1 cells and MT cells on the other hand, are almost all direction selective ~\citep*{mava1983b} and hence are mostly bandpass temporal frequency filters. We firmly believe this classification is not arbitrary, but instead is a direct manifestation of the particular spatiotemporal structure of the V1 to MT motion processing stream. Consequently, the receptive field model must reflect this increased tendency for bandpass-ness with ascension up the hierarchy.  In other words, the representation scheme must be one that is inherently more likely to deliver bandpass-ness to a more specialized cell (such as in layer 4B or MT) and lowpass-ness to a less specialized cell (such as in layer 4A, 4C$\alpha$, or 4C$\beta$). However, none of the existing receptive field models reflect this underlying spatiotemporal structure. In contrast, as will be seen in Section~(\ref{sec:Gabor-Einstein Wavelet}) below, the Gabor-Einstein wavelet's wave carrier is a sinc function which directly confers the aforementioned salient property. Individual Gabor-Einstein basis elements are lowpass temporal frequency filters; and bandpass temporal frequency filters can only be obtained by combinations of basis elements. A study by DeAngelis and colleagues also corroborates the above. They found that temporally monophasic V1 cells in the cat were almost always low pass temporal frequency filters, while temporally biphasic or multiphasic V1 cells were almost always bandpass temporal frequency filters~\citep*{deoh1993b}. The explanation for this finding is inherent and explicit in the Gabor-Einstein wavelet basis, where bandpass temporal character necessarily results from biphasic or multiphasic combination. All monophasic elements on the other hand, are lowpass temporal frequency filters. However, we will see that according to the model, the converse is not true: i.e. not all lowpass temporal frequency filters are monophasic, and not all multiphasic combinations yield bandpass temporal frequency filters.

\subsection{Analogy to Special Relativity}

Perceptual and electrophysiological studies reveal intriguing similarities between how space and time are mixed in the visual cortex and how they are mixed in the special theory of relativity (STR). For instance, within certain limits, binocular neurons cannot distinguish a temporal delay from a spatial difference. As a result, inducing a monocular time delay by placing a neutral density filter over one eye but not the other, causes a pendulum swinging in a 2D planar space to be perceived as swinging in 3D depth space. This is known as the Pulfrich effect. Also, most neurons in the V1 to MT stream are maximally excited only by stimuli moving at that neuron's preferred speed. They are \textit{speed tuned}. In this subsection, we briefly review and summarize the special theory of relativity. We identify the stroboscopic pulfrich effect and speed tuning as cortical analogues of STR's joint encoding of space and time. And finally, we explain how STR is used in (and motivates) the design of the Gabor-Einstein wavelet. 

\subsubsection{Review of Special Relativity}

In 1905, Albert Einstein proposed the special theory of relativity~\citep{ei1905}. He was motivated by one thing: a firm belief that Maxwell's equations of electromagnetism are \textit{laws of nature}. Maxwell's equations are a classical description of light~\citep{ma1861a,ma1861b,ma1862a,ma1862b}. According to Maxwell, light is the propagation of electric and magnetic waves interweaving in a particular way perpendicular to each other. The speed of propagation in a vacuum is the constant $c=299,792,458$ m/s. The other governing constraint on STR is the \textit{relativity principle}. It has been around and generally accepted since the time of Galileo, and it states that the laws of nature are true and the same for all non-accelerating observers. Einstein believed Maxwell's equations to be laws of nature, and therefore to satisfy the relativity principle. In other words, Einstein believed the speed of light must be the same to all non-accelerating observers regardless of their velocity relative to each other. For the speed of light to be fixed, something(s) had to yield and be unfixed. What had to yield were the components of the definition of speed, i.e. space and time. Space and time had to become functions of relative velocity and of each other, to ensure the speed of light remains constant to all non-accelerating observers. The transformation rule ---of space and time coordinates between two observers moving with constant velocity relative to each other--- is called the Lorentz equations. It specifies how space and time are mixed in STR. More specifically, the Lorentz equations are the transformation rule between the (spacetime) coordinates of two inertial reference frames. For two inertial reference frames, $R$ and $R'$, moving exclusively in the $x$ direction with constant velocity $v$ relative to each other, the Lorentz transformation is given by, 

\begin{equation}
\begin{array}{l}
\displaystyle t' =  \gamma\left( t-\frac{v}{c^2} x\right),\\
\displaystyle  x' = \gamma( x-vt),\\
\end{array}
\label{Lorentz_transform}
\end{equation}

where ($t',x'$) are the coordinates of $R'$, ($t,x$) are the coordinates of $R$, $c$ is the speed of light, and $\gamma$ is the Lorentz factor and is given by, 

\begin{equation}
 \gamma = \frac{1}{\sqrt{1-\frac{v^2}{c^2}}}.
\end{equation}

We demonstrate below that the stroboscopic pulfrich effect is analogous to STR in one respect, and speed tuning of neurons is analogous to STR in another respect.

\subsubsection{The Stroboscopic Pulfrich Effect}

In the classical pulfrich effect, a monocular temporal delay results in a perception of depth in pendular swing~\citep{pu1922}. It has a simple geometric explanation. By the time the temporally delayed eye ``sees'' the pendulum at a given retinal position, the pendulum has advanced to a further position in its trajectory, hence the visual cortex always receives images that are spatially offset between the eyes. The spatial disparity results in stereoscopic depth perception. This form of pulfrich phenomena is termed \textit{classical} to distinguish it from the stroboscopic pulfrich effect. In the stroboscopic pulfrich effect, the pendulum is not a continuously moving bulb, but instead is a strobe light which samples the trajectory of the classical pendulum in time and space~\citep*{le1970,mo1976,buro1979,mo1979}. The stroboscopic pulfrich effect cannot be explained by the simple geometric illustration that explains the classical pulfrich effect. This is because depth is still perceived in sequences involving a temporal delay in flash between eyes, but particularly lacking an interocular spatial disparity in the flash~\citep{buro1979}.

\begin{figure}[htp]

  \centering
  \subfloat[]{\label{fig:pulfrich2d}\includegraphics[width=65mm]
   {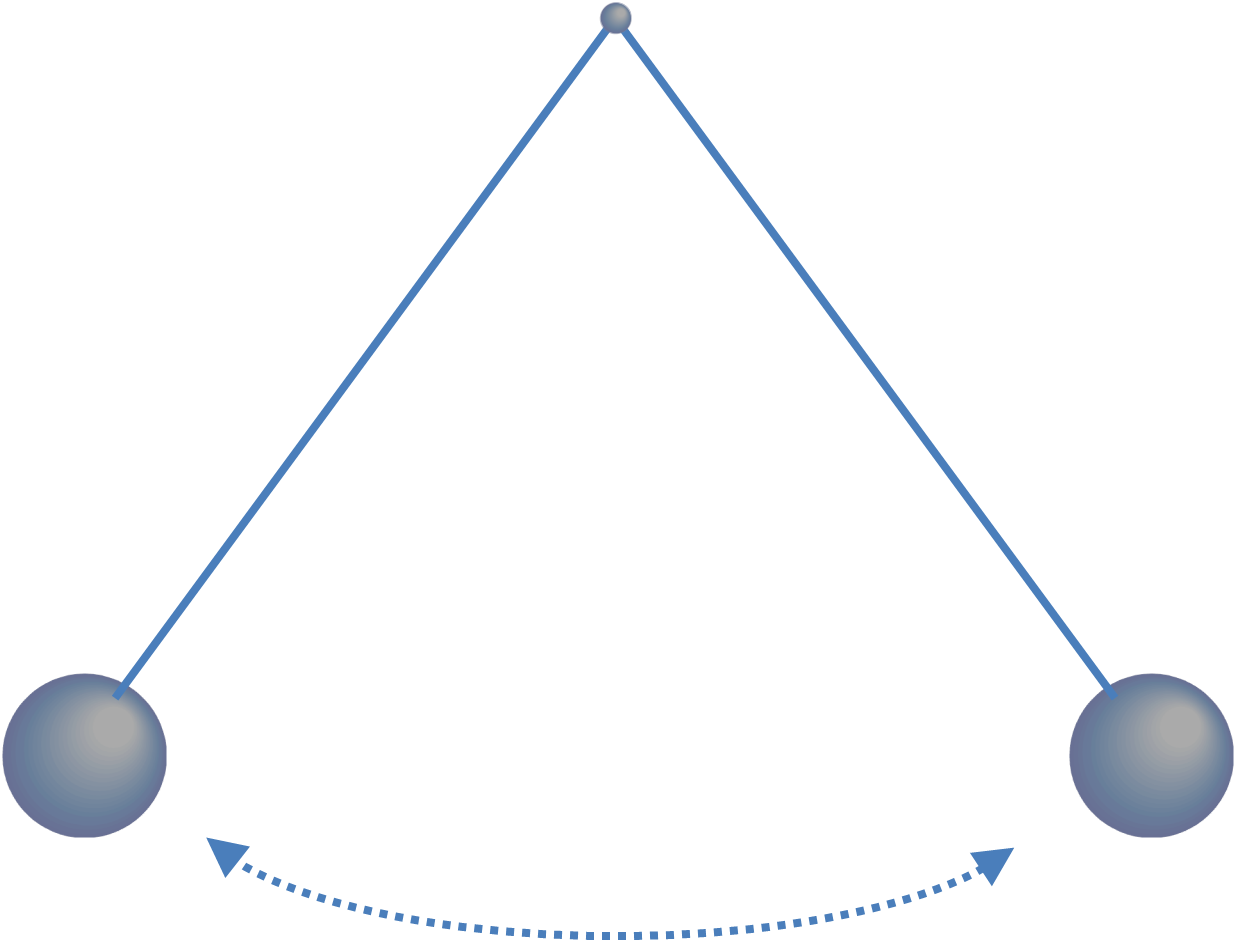}} \\
  \subfloat[]{\label{fig:pulfrich3d}\includegraphics[width=63mm]
   {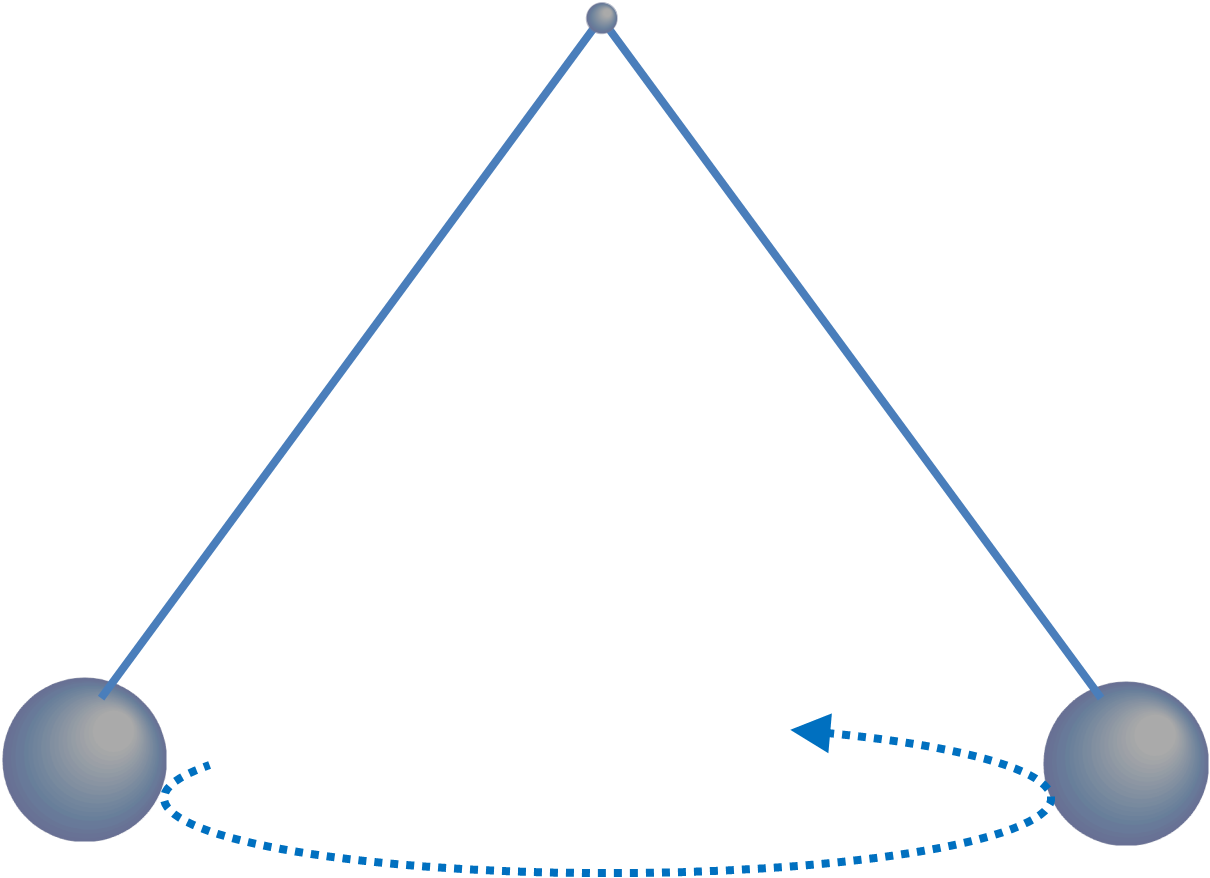}}

\caption{The pulfrich effect: Top figure, (a), shows planar 2D pendular swing. Bottom figure, (b), shows perception of 3D pendular swing (depth) resulting from interocular temporal delay.}
\label{fig:pulfrich}
 \end{figure}

In stereoscopic depth perception, a planar (x,y) difference in retinal image position ($\Delta x$) \textit{transforms into} (is perceived as) a displacement along the $z$ direction. In the stroboscopic pulfrich effect, a temporal difference ($\Delta t$) between eyes in retino-cortical transmission of image signal, results in a planar pendulum trajectory being perceived as (transforming into) depth. In STR, the transformation is between two inertial reference frames moving relative to each other. By analogy, we assert that for the stroboscopic pulfrich effect, there is a transformation between the interocular retinal space ($x_r,t_r$) and the perceptual depth space ($x_z,t_z$). We identify the variables ($x_r,t_r$) of the interocular space as the difference between eyes of the corresponding variables in the retino-cortical space i.e. $x_r$ is the difference between the left and right eye retinal positions of an image; and $t_r$ is the difference between the right and left eyes in retino-cortical image transmission times. The $(x_r,t_r)\rightarrow(x_z,t_z)$ transformation may take the form $x_z=\alpha(x_r\pm \tilde{v}t_r)$, were $\tilde{v}$ is proportional to pendulum velocity and $\alpha$ is a function of $\tilde{v}$. Note the similarity between $x_z$ and the equation for $x'$ in the Lorentz transformation, $x'=\gamma(x-vt)$. This analogy suggests that similar mathematical structures underlie STR and the joint encoding of space and time in the visual cortex. The stroboscopic pulfrich effect is not by itself direct evidence of joint encoding of spatial and temporal disparities ~\citep*{recu2005,recu2005b,recu2005c}, however direct electrophysiological evidence does exist for such linkage in cortical encoding~\citep*{capa1989,anoh2001,pabo2003}. This mathematical similarity influences our design of the Gabor-Einstein wavelet in a particular way which we describe in Section~(\ref{Gabor_Einstein_Design}) below. But first, we discuss another neuronal phenomenon, speed tuning, which is analogous to STR in a certain physical sense. 

\subsubsection{Speed Tuning}

Most motion processing stream neurons in V1 or area MT are speed tuned. They have a preferred speed, which is the speed of moving stimuli to which they maximally respond. Other speeds also elicit a response, but with a lower rate of action potential. When presented moving sine wave grating stimuli, truly speed tuned neurons respond maximally to their preferred speed independent of the spatial or temporal frequency of the stimuli. The velocity, $\xi$, of a moving wave is,

\begin{equation}
 \xi=\frac{\omega}{u}, 
\end{equation}

where $\omega$ is temporal frequency and $u$ is spatial frequency. In higher dimensions,

\begin{equation}
 \omega = u\xi_x + v\xi_y,
\end{equation}

where $u$ is the spatial frequency in the $x$ direction, $v$ is the spatial frequency in the $y$ direction, $\xi_x$ is the $x$ component of velocity, and $\xi_y$ is the $y$ component of velocity. From both equations above, it is clear that for a speed tuned neuron, a change in the spatial frequency of the stimuli would necessary require a complementary change in the temporal frequency. Hence a broad range of stimuli with widely varying spatial and/or temporal frequencies can maximally excite the neuron, so long as the above equations are satisfied for the neuron's preferred speed $\xi_0$. This is analogous to how the special theory of relativity is motivated by fixing the speed of light. In the case of STR the variables which necessarily change in complementary fashion are the space and time coordinates as described in the Lorentz equations.

\subsection{Design of the Gabor-Einstein Wavelet}
\label{Gabor_Einstein_Design}
We constrained the Gabor-Einstein model by requiring the following properties: 

\begin{enumerate}
 \item The minimum possible number of parameters
\item Relativistic-invariance of the wave carrier
\end{enumerate}

\subsubsection{Minimal Number of Parameters}

In addition to orientation in the spatiotemporal frequency domain, localization in the space, time, spatial frequency, and temporal frequency domains are the essential features of the V1 to MT neuron receptive field. These are the most basic attributes of the receptive field. As such, they correspond with the minimum number of model parameters needed. Specifically, speed tuning, the spatial and the temporal localization envelopes, the amplitude modulation parameter, and the spatial and temporal frequencies must all be represented. If one adopts the notion of a gaussian spatial localizer~\citep{ma1980,da1985}, the minimum number of parameters add up to eight: four for the spatial envelope and four for the wave carrier. The four necessary spatial envelope parameters are: amplitude factor, $A$; variance in the $x$ direction, $\sigma_x$; variance in the $y$ direction, $\sigma_y$; and spatial orientation of envelope, $\theta_e$. The four necessary wave carrier parameters are: temporal frequency, $\omega$; spatial frequency in the x direction, $u$; spatial frequency in the $y$ direction $v$, and the phase, $\phi$. The orientation of the wave carrier $\theta_s=\tan^{-1}(v/u)$ is not an independent parameter. We use it in computer simulations below only for convenience. The essential feature that is not yet explicitly accounted for in the above parameter tally is the temporal envelope of the receptive field. This can be implemented either through the gaussian envelope or through the wave carrier, and in either case may conceivably involve an additional parameter. The next design criteria, \textit{relativistic-invariance of the wave carrier}, resolves this ambiguity without adding another parameter.

\subsubsection{Relativistic-invariance of the Wave Carrier}

Given the analogies of spatiotemporal mixing in the visual cortex to spatiotemporal mixing in the special theory of relativity, it is reasonable to conjecture that their endpoints are within proximity of each other. Einstein eventually sought a lexicon in which physical laws are expressed the same way in any two inertial reference frames. That lexicon is called relativistic-invariance (or Lorentz-invariance). By analogy, we ultimately seek a lexicon in which physical laws can be expressed the same way in both the interocular and perceptual spaces. Lorentz-invariance of physical law is the specification (or endpoint) arising out of STR. Hence, requiring Lorentz-invariance of the receptive field's wave carrier is desirable. Moreover, this immediately resolves the above ambiguity regarding implementation of the temporal envelope profile. The temporal envelope must be implemented via the wave carrier for relativistic invariance to hold. And since the receptive field amplitude eventually falls, the sinc function emerges as a natural descriptor. Furthermore, the sinc function has the distinct advantage of not introducing any additional parameters.

\subsection{Summary}
The way space and time are mixed in the visual cortex bears resemblance to the way they are mixed in the special theory of relativity. We demonstrated that the stroboscopic pulfrich effect and speed tuning  are cortical analogues of the spacetime mixing mechanics of STR. In STR, the mixture of space and time is described by the Lorentz transform which relates the spacetime coordinates of inertial reference frames moving with constant velocity relative to each other. In the striate and extrastriate motion processing stream, the analogous transformation is between the 3D interocular space (x,y planar inter-retinal space + inter-retinocortical time) and the 3D perceptual space (x,y,z 3D space). Though on the surface, STR and cortical spacetime encoding appear to be unrelated processes, we conjecture and have partly shown a shared underlying mathematical structure. This structure can be exploited to deepen our understanding of cortical encoding of motion, depth, and more fundamentally, the joint encoding of space and time in the visual cortex. Here, we proceeded by requiring relativistic-invariance of the receptive field's wave carrier. We simultaneously required that the model be constrained to the minimum possible number of parameters. Under these constraints, the sinc function with energy-momentum relation as argument emerged as a natural descriptor of the receptive field's wave carrier. Furthermore, the Gabor-Einstein wavelet explains a number of salient physiological attributes of the V1 to MT spectrum. Chief amongst these being the distribution of bandpass to lowpass temporal frequency filter distribution profile; which we postulate is a fundamental manifestation of the way space and time are mixed in the visual cortex.

The remainder of this paper is organized as follows: Section~(\ref{sec:Gabor-Einstein Wavelet}) introduces the Gabor-Einstein wavelet. Section~(\ref{sec:simulations}) presents computer simulations. Section~(\ref{sec:discussion}) is the discussion. It includes the following subsections: Subsection~(\ref{subsec:model_framework}) introduces a simple framework for classifying and naming the components and hierarchical levels of receptive field models. Subsection~(\ref{subsec:model_framework}) also identifies the Gabor-Einstein's place within this larger framework of receptive field modeling. Subsection~(\ref{subsec:related_work}) briefly discusses related work. Subsection~(\ref{subsec:higher_order}) discusses some phenomena which cannot be explained by models such as the native Gabor-Einstein wavelet in isolation, i.e. models of single neuron receptive fields early in the visual pathway. Specifically, it discusses some higher order phenomena and non-linearities which necessarily arise from neuronal population and network interactions. Section~(\ref{sec:conclusion}) concludes the paper.

\section{The Gabor-Einstein Wavelet}
\label{sec:Gabor-Einstein Wavelet}

In this section, we present the Gabor-Einstein wavelet. It has the following properties:

\begin{itemize}
\item Like the Gabor function, it is a product of a gaussian envelope and a sinusoidal wave carrier.
\item It differs from the Gabor function in that its wave carrier is a relativistically-invariant sinc function whose argument is the energy-momentum relation.
\item Its gaussian envelope contains only spatial arguments. i.e. its spatial envelope is not time-dependent.
\item It has the minimum possible number of parameters.
\item Its fourier transform is the product of a mixed frequency gaussian and a temporal frequency step function.
\item Like the Gabor function, it generates a quasi-orthogonal basis.
\end{itemize}

We define the Gabor-Einstein wavelet as follows,

\begin{equation}
\label{Lorentz_Gabor_l}
 G(t,x,y) = A~\exp\left(-\frac{(x-x_0)^2}{2\sigma^2_x}-\frac{(y-y_0)^2}{2\sigma^2_y}\right)\mbox{sinc}(\omega_0 t -u_0 x-v_0 y+\phi),
\end{equation}

where the constant multiplier $A$ is amplitude; $\sigma_x$ and $\sigma_y$ are the gaussian variances in the x and y directions; $x_0$ and $y_0$ are the respective $x$ and $y$ coordinates of the gaussian center; $\omega_0$, $u_0$, and $v_0$ are the frequencies of the wave carrier in the $t$, $x$, and $y$ directions respectively; $\phi$ is the sinusoid phase; and the \textit{sinc} function is defined as,

\begin{equation}
 \mbox{sinc}(x) = \frac{\sin(x)}{x}.
\end{equation}

We have rotated our coordinates by an angle $\theta$ from a reference state $(x',y')$ to a state $(x,y)_\theta$ which we denote $(x,y)$ for notational simplicity. The transformation,

\begin{equation}
 (x',y')~~~\longrightarrow ~~~(x,y)_\theta :=(x,y),
\label{Eqn:envelope_axes_rotate_1}
\end{equation}
is given by,

\begin{equation}
 \vectwo{x}{y} = \left( \begin{array}{cc}
\cos (\theta) & -\sin (\theta)\\
\sin (\theta) & ~~\cos (\theta)\\
\end{array} \right)\vectwo{x'}{y'}.
\label{Eqn:envelope_axes_rotate_2}
\end{equation}

We proceed here with the following instance of the Gabor-Einstein wavelet,

\begin{equation}
\label{Gabor-Einstein}
 G(t,x,y) = \exp\left(-\frac{x^2}{2\sigma^2_x}-\frac{y^2}{2\sigma^2_y}\right)\mbox{sinc}(t -u_0 x-v_0 y), 
\end{equation}

where we have set $\omega_0$ equal to one. We can always do so for one reference neuron by simply defining the unit of time as $1/\omega_0$, the duration the neuron's frame cycle. As we will see, this value, $1/\omega_0$, is the shortest frame duration to which the neuron can respond. In the above equation, we have also set $A=1$, and $\theta=x_0=y_0=\phi=0$.

The fourier transform is as follows,

\begin{equation}
\label{Gabor-Einstein-Fourier}
 H(\omega,u,v)=\frac{\sigma_x\sigma_yA}{2}\sqrt{\frac{\pi}{2}}\exp\left(-\frac{\sigma_x^2(u+u_0\omega)^2}{2}-\frac{\sigma_y^2(v+v_0\omega)^2}{2}\right)L(\omega), 
\end{equation}

where $L(\omega)$ is given by,

\begin{equation}
\label{eqn:sign}
 L(\omega) = \mbox{sign}(1-\omega) + \mbox{sign}(1+\omega)
\end{equation}

and the \textit{sign} function is defined as,

\begin{equation}
 \mbox{sign}(x) = \begin{cases}
            ~~1 & \mbox{if}~x>0,\\
~~0 & \mbox{if}~x=0, ~\mbox{and}\\
-1 & \mbox{if}~ x<0.
           \end{cases}
\end{equation}

The above fourier transform was obtained using Mathematica symbolic software. The fourier transform of the general case will likely be challenging to obtain analytically either by hand or symbolic software. Hence, we anticipate numerical methods may have an important role to play.

The maximum magnitude of the above response function, Equation~(\ref{Gabor-Einstein-Fourier}), is attained where the argument of the exponent is zero, i.e. where,

\begin{equation}
\label{F_max}
\sigma_x^2(u+u_0\omega)^2+\sigma_y^2(v+v_0\omega)^2=0.
\end{equation}

The neuron's preferred spatial frequency, $f_0$, is dependent on the temporal frequency, $\omega$, of the stimulus, and is given by the above bivariate quadratic equation's solution, 

\begin{equation}
\label{Eqn:f_0}
 f_0(\omega)=(-u_o\omega,-v_0\omega).
\end{equation}

The location where the response maximum is attained is a parametrized curve in 3D frequency space. It is given by,

\begin{equation}
 R(\omega)=(\omega, -u_0\omega,-v_0\omega).
\end{equation}

Although the location in $(u,v)$ space where the neuron's maximum response is attained depends on $\omega$, the magnitude of the maximum response is itself a constant. It is given by,

\begin{equation}
 R_{max} = |H(\omega,f_0(\omega))|= \frac{\sigma_x\sigma_yA}{2}\sqrt{\frac{\pi}{2}}.
\end{equation}

The half magnitude response is attained at values of $(u,v)$ satisfying,

\begin{equation}
 \exp\left(-\frac{\sigma_x^2(u+u_0\omega)^2}{2}-\frac{\sigma_y^2(v+v_0\omega)^2}{2}\right) = \frac{1}{2}.
\end{equation}

Taking the natural logarithm of the above equation yields,
\begin{equation}
\label{half_max}
 \frac{\sigma_x^2(u+u_0\omega)^2}{2~\mbox{ln}(2)}+\frac{\sigma_y^2(v+v_0\omega)^2}{2~\mbox{ln}(2)} = 1.
\end{equation}

Defining $\sigma_x^{-1}\sqrt{2~\mbox{ln}(2)}\rightarrow a$ and $\sigma_y^{-1}\sqrt{2~\mbox{ln}(2)} \rightarrow b$, Equation~(\ref{half_max}) becomes,

\begin{equation}
\label{half_max2}
  \frac{(u+u_0\omega)^2}{a^2}+\frac{(v+v_0\omega)^2}{b^2} = 1.
\end{equation}

In the above form, one readily recognizes this as the ellipse centered at $(-u_0\omega,-v_0\omega)$, whose principal axes radii are $a$ and $b$ in the $x$ and $y$ directions respectively. The long axis in the frequency domain is the short axis in the spatial domain and vice versa. Without loss of generality, we can assume that prior to rotation of the spatial axes by angle $\theta$, the long axis of the receptive field envelope is parallel to the x axis. Then the orientation of the on-off bands, i.e. planes of the spatial wave, are aligned parallel to the long axis of the envelope when the rotation angle, $\theta$, is related to the polar angle $\mu$ by the relationship,

\begin{equation}
\label{parallel}
 \theta= \mu.
\end{equation}

On the other hand, the on-off bands are perpendicular to the long axis of the envelope when,

\begin{equation}
\label{perpendicular}
 \mu=\theta+\frac{\pi}{2}.
\end{equation}

In the case of Equation~(\ref{parallel}), the half magnitude frequency bandwidth is readily seen to equal $2a$, the length of the short axis, while in the case of Equation~(\ref{perpendicular}), the half magnitude frequency bandwidth equals $2b$, the length of the short axis. The cases of skewed alignment take on values between $2a$ and $2b$ and are also computable from the geometry. Unlike the preferred spatial frequency, the half-magnitude frequency bandwidth is not dependent on the temporal frequency of the stimulus.

 In summary, the Gabor-Einstein wavelet spatiotemporal receptive field model predicts that the magnitude of a V1 to MT neuron's preferred spatial frequency is linearly dependent on the temporal frequency of the stimulus as shown in Equation~(\ref{Eqn:f_0}). In the next section, we present some receptive field simulations using the Gabor-Einstein wavelet.

\begin{figure}[htp]
                                                                         
  \centering
  \subfloat[]{\label{fig:Gaussian}\includegraphics[width=60mm]{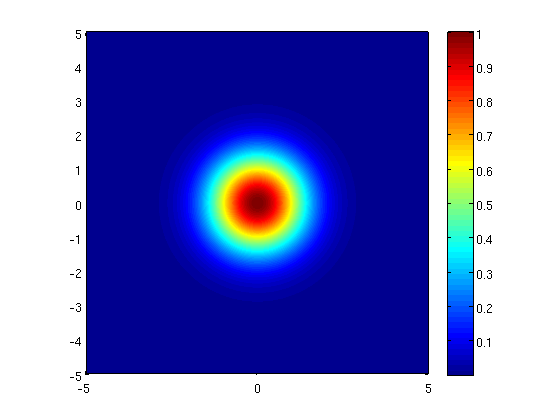}}   
  \subfloat[]{\label{fig:Gaussian2}\includegraphics[width=60mm]{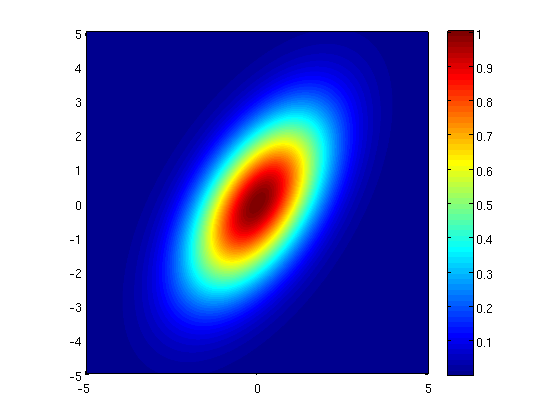}}   
  \\
  \subfloat[]{\label{fig:Gaussian4}\includegraphics[width=60mm]{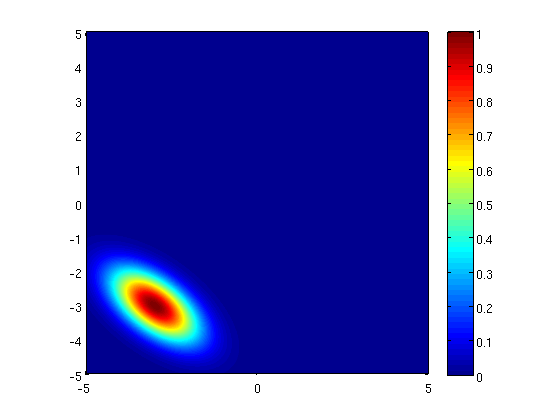}}    
  \subfloat[]{\label{fig:Gaussian5}\includegraphics[width=60mm]{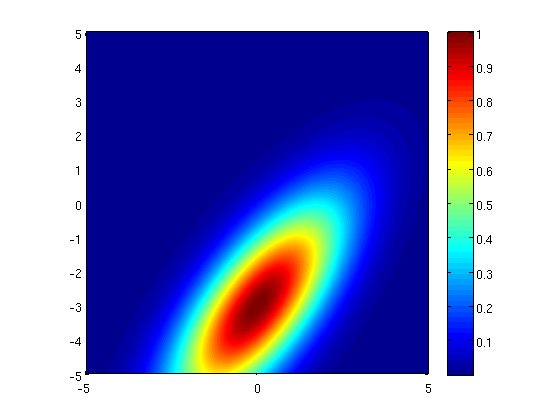}}    

 \label{figur_gaussian}\caption{Receptive field spatial envelopes. The following parameters were used in the above gaussians. In Fig(\ref{fig:gaussians}a), $\theta=1$, $\sigma=(1,1)$, and $\mbox{ctr}=(0,0)$; In Fig(\ref{fig:gaussians}b) $\theta=1$, $\sigma=(2,1)$, and $\mbox{ctr}=(0,0)$; In Fig(\ref{fig:gaussians}c) $\theta=2.5$, $\sigma=(1.0,0.5)$, and $\mbox{ctr}=(-3,-3)$; In Fig(\ref{fig:gaussians}d) $\theta=2.5$, $\sigma=(1.0,2.5)$, and $\mbox{ctr}=(0,-3)$.}
\label{fig:gaussians}
 \end{figure}

\section{Computer Simulations}
\label{sec:simulations}

In this section, we present computer simulations demonstrating the Gabor-Einstein wavelet's properties. We start with simulations which show the model's adherence to basic physiological form. We then present simulations which demonstrate how the Gabor-Einstein model innately represents the temporal frequency filtering property distribution along the V1 to MT neuronal hierarchy.

The following notations are used in the figure captions: $\theta_e$ is the angle of rotation of the axes of the gaussian envelope as described in Equations~(\ref{Eqn:envelope_axes_rotate_1}) and (\ref{Eqn:envelope_axes_rotate_2}). $\theta_s$ is the angle of rotation of the axes of the wave carrier. ctr is the 2-component vector consisting of the $x$ and $y$ coordinates of the gaussian envelope center respectively. $\sigma=(\sigma_x,\sigma_y)$ is the 2-component vector consisting of the gaussian variances of the envelope in the $x$ and $y$ directions respectively. $\phi$ is phase of the wave carrier. $t$ is time. And $\hat{\omega}=(\omega_0,u_0,v_0)$ is the 3-component vector consisting of the temporal, x-spatial, and y-spatial frequencies of the wave carrier respectively. For succinctness, $\hat{\omega}=1$, for instance, is taken to be equivalent to $\hat{\omega}=(1,1,1)$. The same short-hand notation applies to the other multi-component vectors.

\subsection{Basic Physiological Form}

\begin{figure}[htp]
                                                                         
  \centering
  \subfloat[$t=0$, $\hat{\omega}=(1,1,1)$]{\label{fig:sinc}\includegraphics[width=60mm]{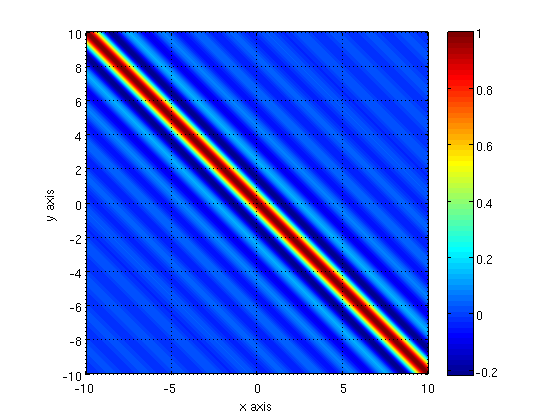}}   
  \subfloat[$t=4$, $\hat{\omega}=(1,1,1)$]{\label{fig:sinc2}\includegraphics[width=60mm]{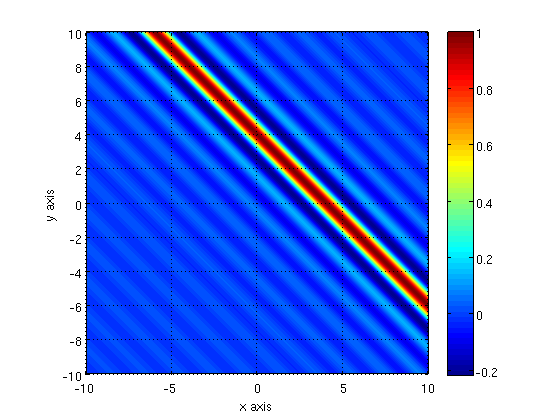}}   
  \\
  \subfloat[$t=12$, $\hat{\omega}=(1,1,1)$]{\label{fig:sinc4}\includegraphics[width=60mm]{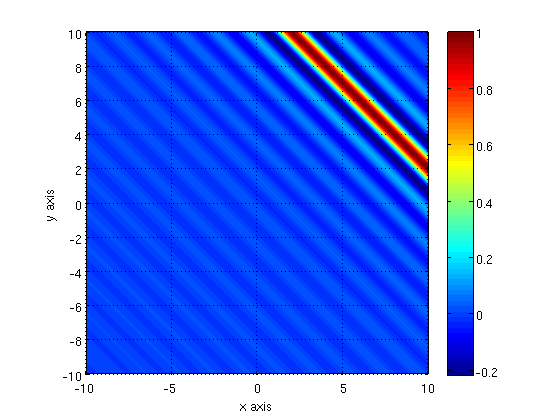}}    
  \subfloat[$t=0$, $\hat{\omega}=(1,2,1)$]{\label{fig:sinc_wx2}\includegraphics[width=60mm]{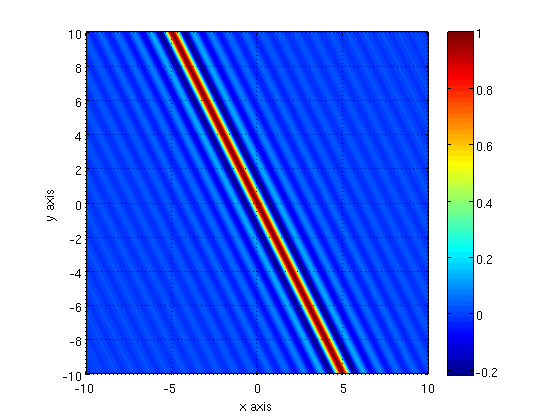}}    

 \label{figur_sinc}\caption{Receptive field wave carrier. 3D sinc waves model the wave carrier. The following parameters were common to each simulation in the figure: $\theta_s=0$, $\phi=0$, $\mbox{ctr}=(0,0)$.}
\label{fig:wavecarriers}
 \end{figure}

\begin{figure}[htp]
  \centering
  \subfloat[$\mbox{ctr}=(0,0)$]{\label{fig:GE000}\includegraphics[width=60mm]{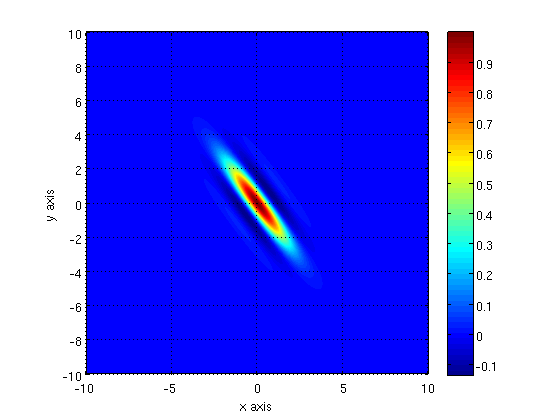}}    
  \subfloat[$\mbox{ctr}=(0.25,0.25)$]{\label{fig:GE025}\includegraphics[width=60mm]{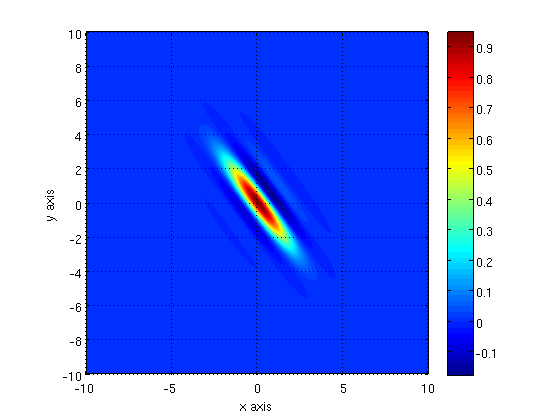}}    
  \\
  \subfloat[$\mbox{ctr}=(0.50,0.50)$]{\label{fig:GE050}\includegraphics[width=60mm]{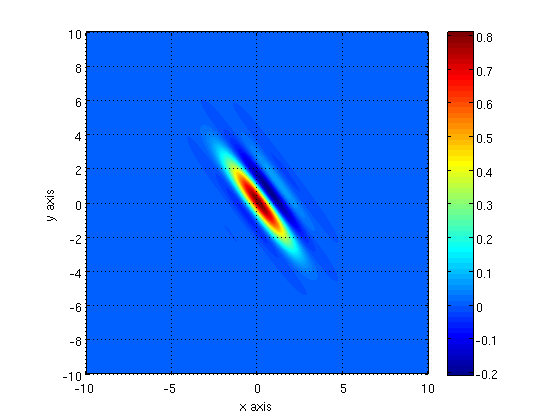}}    
  \subfloat[$\mbox{ctr}=(1,1)$]{\label{fig:GE100}\includegraphics[width=60mm]{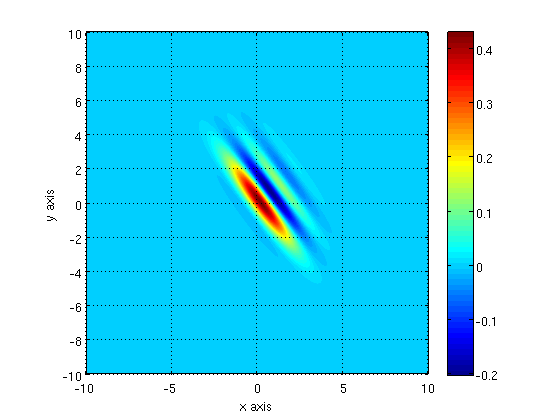}}    
   \\
  \subfloat[$\mbox{ctr}=(1.50,1.50)$]{\label{fig:GE150}\includegraphics[width=60mm]{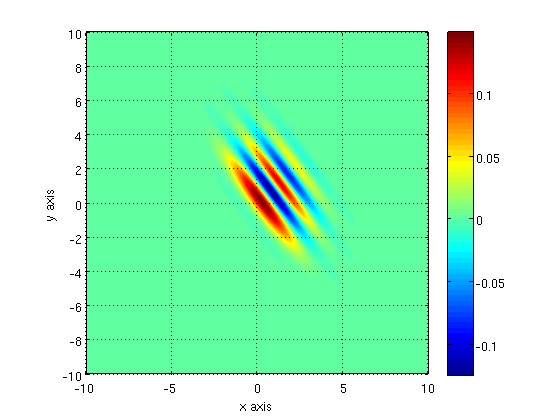}}    
  \subfloat[$\mbox{ctr}=(5.00,5.00)$]{\label{fig:GE500}\includegraphics[width=60mm]{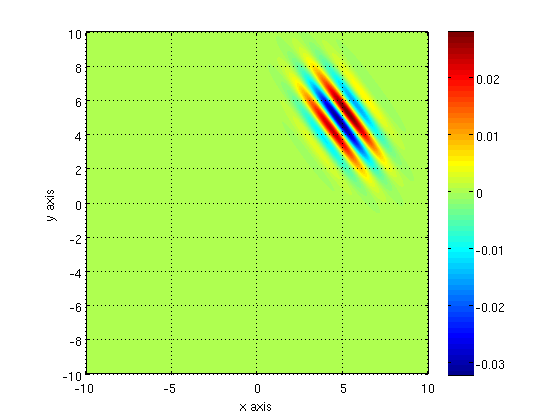}}    

 \label{figur_GE_space}\caption{The Gabor-Einstein wavelet. The following parameters were common to each simulation in the figure: $t=0$, $\hat{\omega}=1$, $\theta_e=10$, $\theta_s=3$, $\phi=0$, $\sigma=(1,2)$.}

\label{fig:gabor-einstein-center}
 \end{figure}

\begin{figure}[htp]
  \centering
  \subfloat[$\mbox{ctr}=(0,0)$]{\label{fig:GE3D000}\includegraphics[width=60mm]{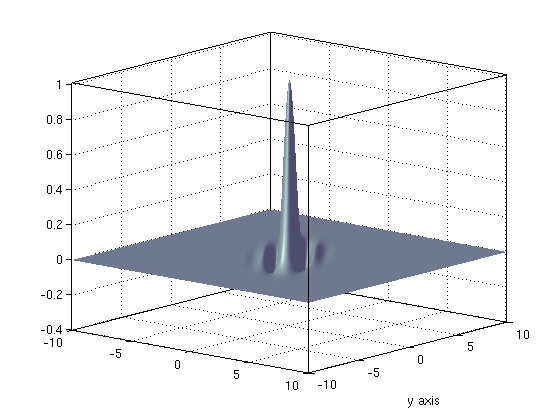}}    
  \subfloat[$\mbox{ctr}=(0.25,0.25)$]{\label{fig:GE3D025}\includegraphics[width=60mm]{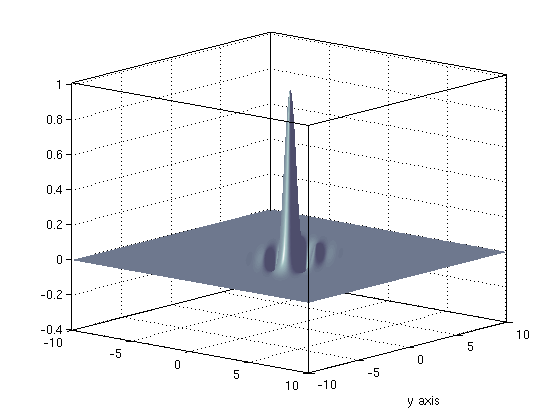}}    
  \\
  \subfloat[$\mbox{ctr}=(0.50,0.50)$]{\label{fig:GE3D050}\includegraphics[width=60mm]{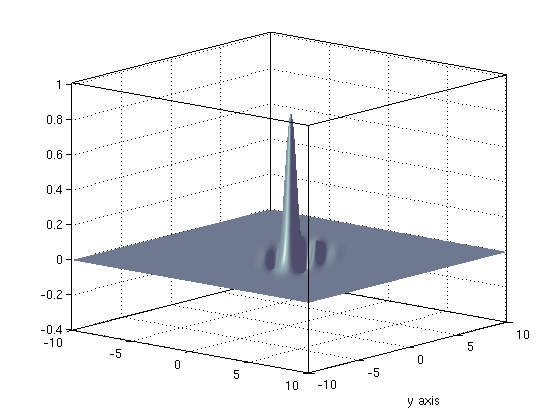}}    
  \subfloat[$\mbox{ctr}=(1,1)$]{\label{fig:GE3D100}\includegraphics[width=60mm]{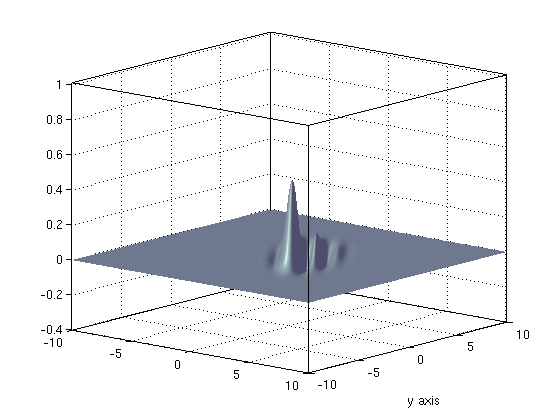}}    
   \\
  \subfloat[$\mbox{ctr}=(1.25,1.25)$]{\label{fig:GE3D125}\includegraphics[width=60mm]{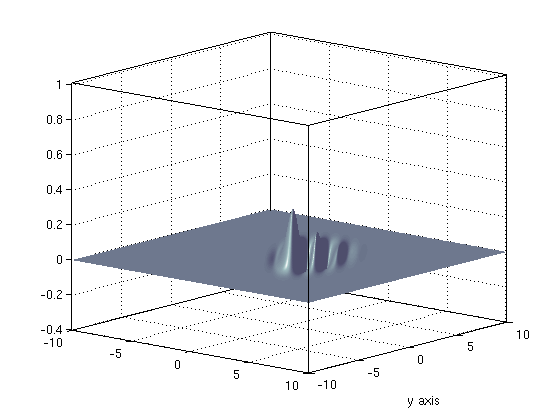}}    
  \subfloat[$\mbox{ctr}=(1.75,1.75)$]{\label{fig:GE3D175}\includegraphics[width=60mm]{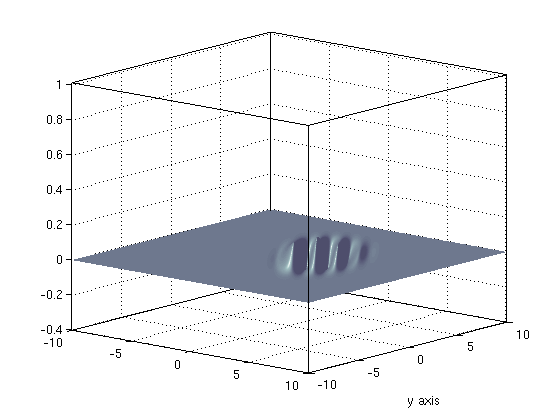}}    

 \label{figur_GE3D_space}\caption{Gabor-Einstein space-modulation. The following parameters were common to each simulation in the figure: $t=0$, $\hat{\omega}=1$, $\theta_e=10$, $\theta_s=3$, $\phi=0$, $\sigma=(1,2)$.}

\label{fig:gabor-einstein-center-3d}
 \end{figure}

\begin{figure}[htp]
  \centering
  \subfloat[$\phi =1$]{\label{figur:1}\includegraphics[width=60mm]{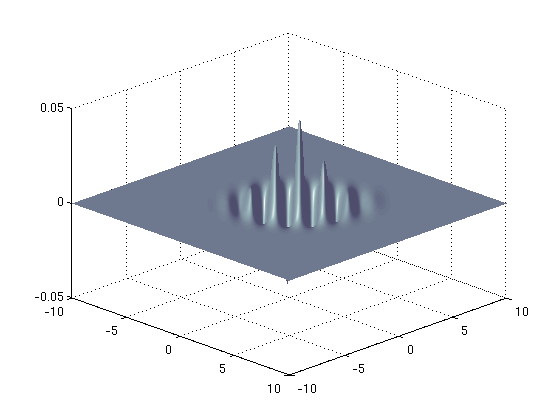}}
  \subfloat[$\phi=2$]{\label{figur:2}\includegraphics[width=60mm]{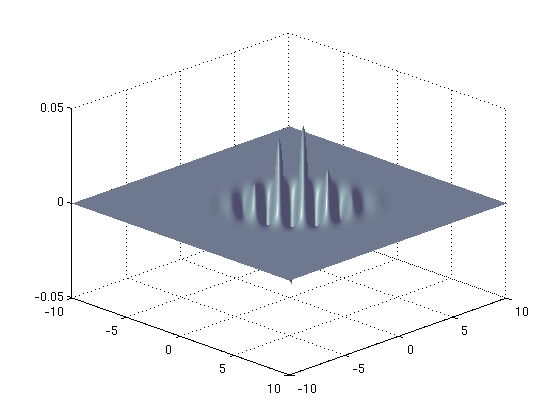}}
  \\
  \subfloat[$\phi=20$]{\label{figur:3}\includegraphics[width=60mm]{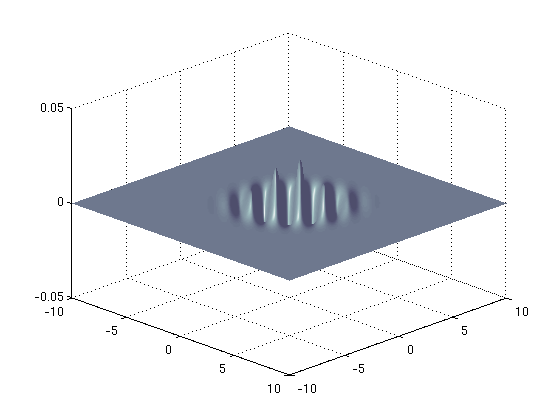}}
  \subfloat[$\phi=40$]{\label{figur:4}\includegraphics[width=60mm]{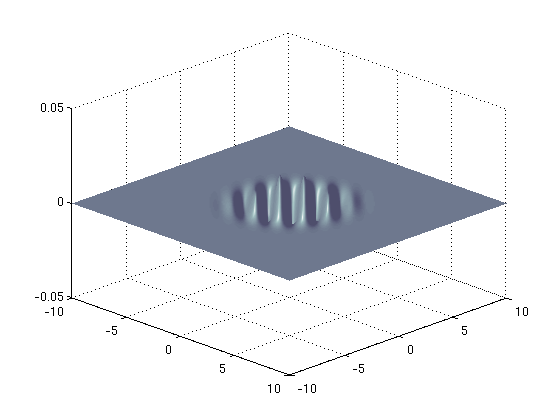}}
 \label{figur_phi}\caption{Gabor-Einstein phase-modulation. The following parameters were common to each simulation in the figure: $t=8$, $\hat{\omega}=2.75$, $\theta=0$, $\mbox{ctr}=(0,0)$, $\hat{\sigma}=1.5$.}

\label{fig:gabor-einstein-spatial-phase}
 \end{figure}

\begin{figure}[htp]

  \centering
  \subfloat[$u_0=-0.25,~v_0=0.075$]{\label{fig:w1}\includegraphics[width=60mm]{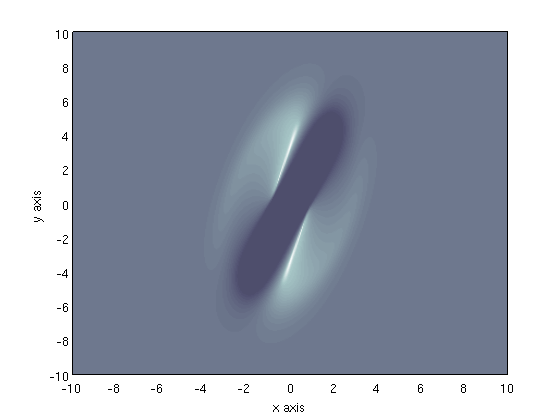}} 
  \subfloat[$u_0=-0.5,~v_0=0.15$]{\label{fig:w2}\includegraphics[width=60mm]{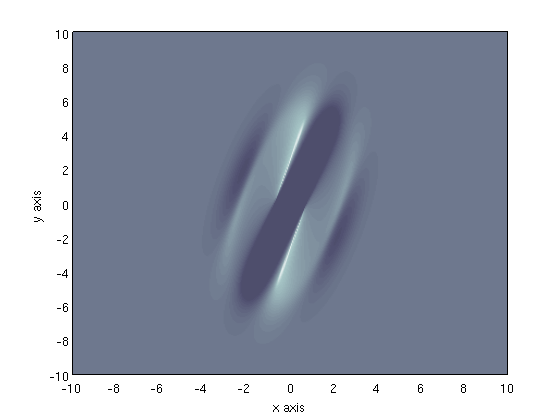}}
  \\
  \subfloat[$u_0=-1,~v_0=0.3$]{\label{fig:w3}\includegraphics[width=60mm]{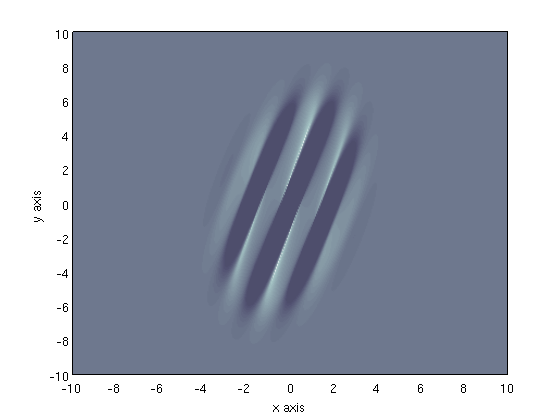}}
  \subfloat[$u_0=-2,~v_0=0.6$]{\label{fig:w4}\includegraphics[width=60mm]{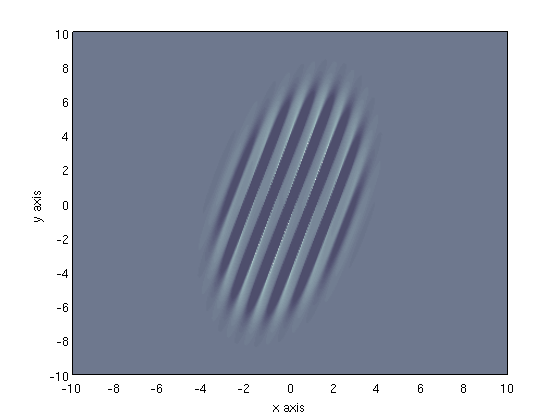}}

 \label{figur_omega}\caption{Gabor-Einstein frequency-modulation. The following parameters were common to each simulation in the figure: $t=10$, $\omega_0=6$, $\theta_s=0$, $\theta_e=-0.25$, $\phi=0$, $\mbox{ctr}=(0,0)$, $\sigma_x=1.0$, $\sigma_y=2.25$.}

\label{fig:gabor-einstein-spatial-freq}
 \end{figure}

 Figure~(\ref{fig:gaussians}) shows receptive field envelopes with varying aspect ratios and centers; Figure~(\ref{fig:wavecarriers}) shows the sinc function wave carriers at varying times $t$ and at one different orientation; Figures~(\ref{fig:gabor-einstein-center}) and (\ref{fig:gabor-einstein-center-3d}) show Gabor-Einstein wavelets with varying envelope centers; Figure~(\ref{fig:gabor-einstein-spatial-freq}) shows Gabor-Einstein wavelets with varying spatial frequencies; and Figure~(\ref{fig:gabor-einstein-spatial-phase}) shows Gabor-Einstein wavelets of varying phase.


In Figure~(\ref{fig:gabor-einstein-center}), one sees the anisometry that is conferred by the sinc function wave carrier. The anisometry increases from Fig (\ref{fig:gabor-einstein-center}a) through (\ref{fig:gabor-einstein-center}d), due to the rapid fall off in the sinc function amplitude outside of the central peak over the two half cycles on each side of zero. In contrast,  Fig (\ref{fig:gabor-einstein-center}e) and Fig (\ref{fig:gabor-einstein-center}f) appear similar to each other in form, and each look more isometric than Fig (\ref{fig:gabor-einstein-center}d). This illustrates the presence of both isometric and anisometric basis elements which differ only in a single parameter (phase, or time, or envelope center). This expressiveness in the basis should allow for economical representation of a visual scene. Of note, the amplitudes of the more isometric elements Figs (\ref{fig:gabor-einstein-center}e) and (\ref{fig:gabor-einstein-center}f) are much smaller than those of the more anisometric preceding elements. If desired, this can be readily modulated by the constant amplitude factor parameter. 

\newpage

Figure~(\ref{fig:gabor-einstein-center-3d}) shows 3-dimensional plots illustrating the above discussed isometry modulation feature of the Gabor-Einstein basis. Figure~(\ref{fig:gabor-einstein-spatial-freq}) shows receptive fields with various spatial frequency tuning. Figure~(\ref{fig:gabor-einstein-spatial-phase}) shows various receptive fields which are phase-shifted relative to each other. For any given spatial location in the receptive field, the peak amplitude of a cycle is phase dependent. This is a feature distinct to the Gabor-Einstein model. And like the anisometry-modulation feature, it confers expressiveness to the basis family. This should allow for efficient representation of natural visual scenes. In the model, time has a similar effect on peak amplitude. And this behavior is in agreement with space-time inseparable receptive field profiles in cat striate cortex, where a time dependent phase drift has been shown~\citep*{deoh1993}.

\clearpage


\clearpage

\begin{figure}[htp]

  \centering
  \subfloat[Sinc, $\omega_0=1$]{\label{fig:sinc_fcn}\includegraphics[width=60mm]
   {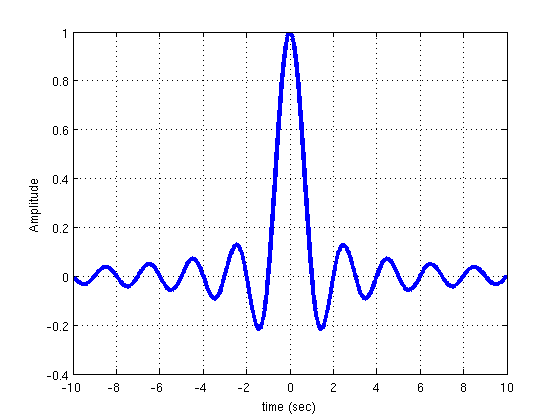}} 
  \subfloat[``\textit{Fourier of (a)}'']{\label{fig:sinc_fourier}\includegraphics[width=60mm]{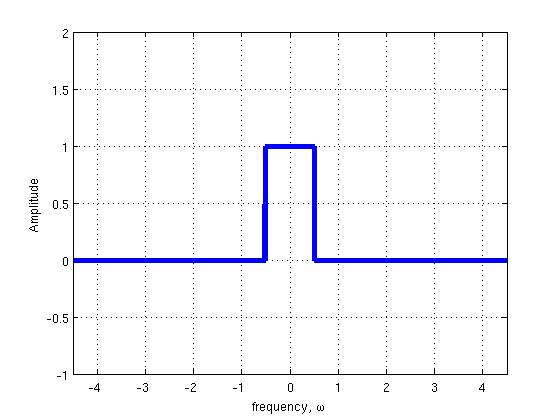}}
\\
  \subfloat[Sinc, $\omega_0=3$]{\label{fig:sinc_fcn_w3}\includegraphics[width=60mm]
   {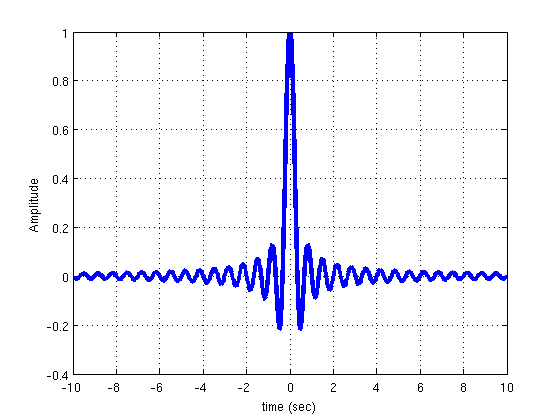}} 
  \subfloat[``\textit{Fourier of (c)}'']{\label{fig:sinc_fourier_w3}\includegraphics[width=60mm]{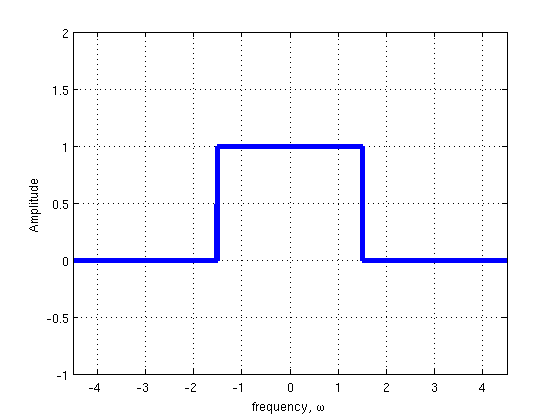}}
\\
  \subfloat[``\textit{(c) minus (a)}'']{\label{fig:sinc_fcn_w3-w1}\includegraphics[width=60mm]
   {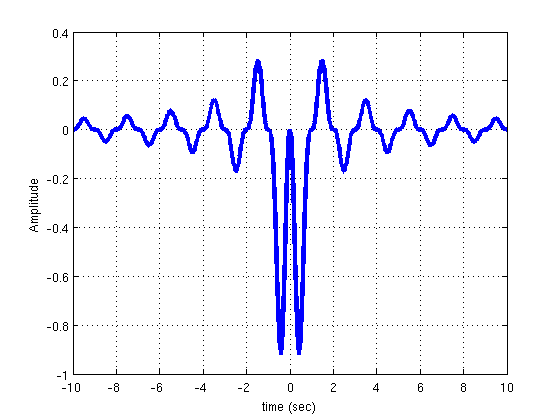}} 
  \subfloat[``\textit{(d) minus (b)}'']{\label{fig:sinc_fourier_w3-w1}\includegraphics[width=60mm]{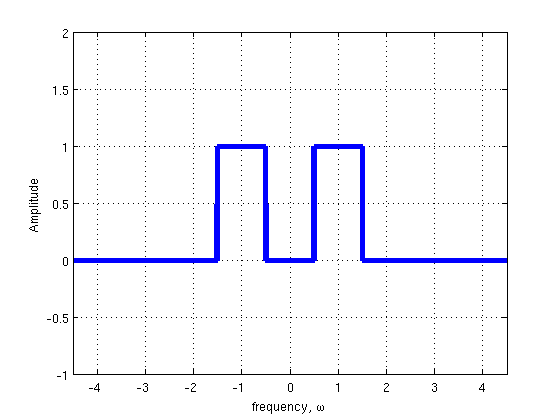}}
 \label{figur_sinc_fcn}\caption{Sinc function and its Fourier transform}

\label{fig:sinc and fourier}
 \end{figure}

\begin{figure}[htp]

  \centering
  \subfloat[]{\label{fig:GEF}\includegraphics[width=60mm]
   {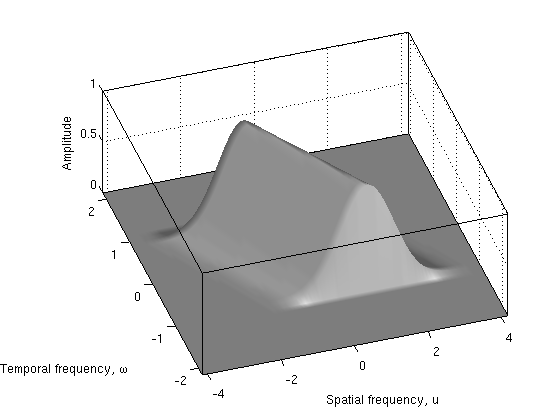}} 
  \subfloat[]{\label{fig:GEF_p}\includegraphics[width=60mm]
   {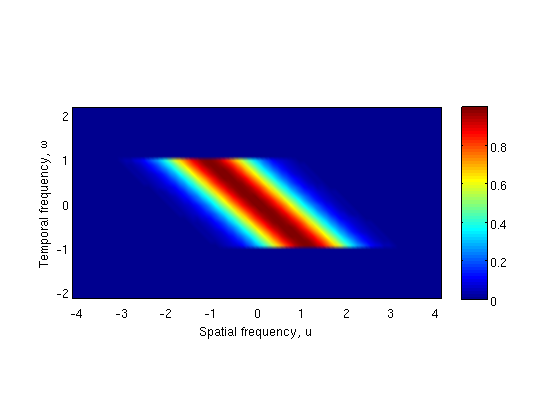}}\\
  \subfloat[]{\label{fig:GEF_xt}\includegraphics[width=60mm]
   {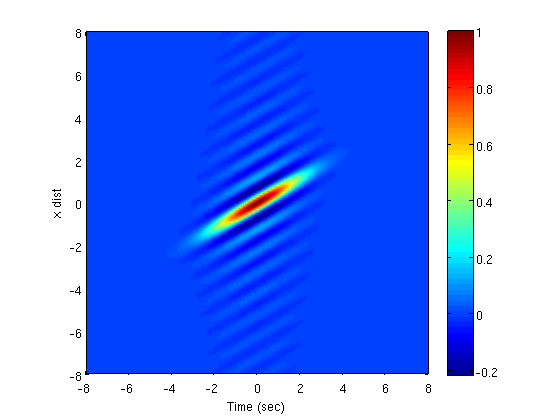}}

\caption{Fourier representation of the 2D (x,t) Gabor-Einstein wavelet. $\omega_0=2,~\sigma=1,~u_0=1.15$. (a) is the surface plot, while (b) is the p colormap. The 3D analogue is more readily imagined as an extension of Figure (b). In particular, Figure (b) is a slice perpendicular to the y-spatial frequency (or ``$v$'') direction of the 3D extension. (c) is the (x,t) domain representation.}
\label{fig:GE_Fourier}
 \end{figure}

\begin{figure}[htp]

  \centering
  \subfloat[$\omega_b=1$]{\label{fig:GEF1}\includegraphics[width=60mm]
   {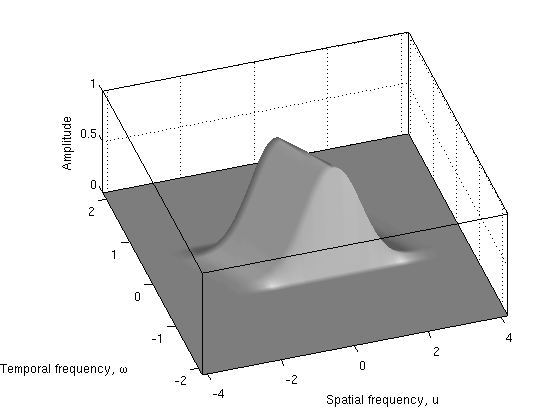}} 
  \subfloat[$\omega_b=1$]{\label{fig:GEF2}\includegraphics[width=60mm]{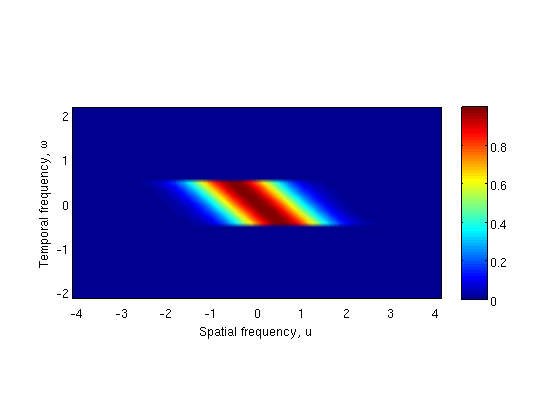}}
\\
  \subfloat[$\omega_b=3$]{\label{fig:GEF3}\includegraphics[width=60mm]
   {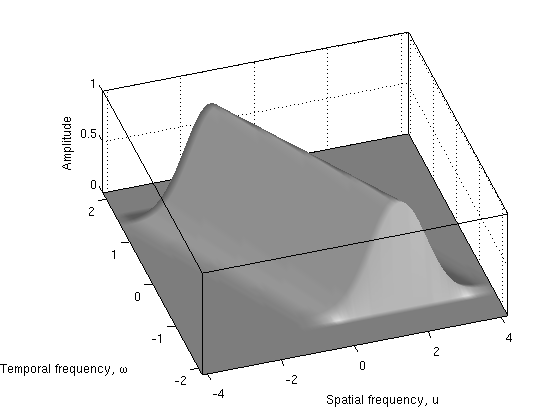}} 
  \subfloat[$\omega_b=3$]{\label{fig:GEF4}\includegraphics[width=60mm]{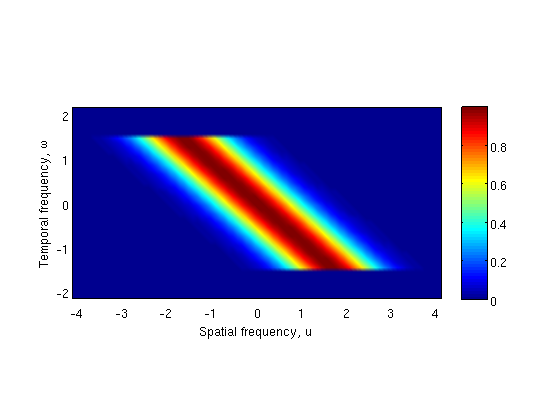}}
\\
  \subfloat[``\textit{(c) minus (a)}'']{\label{fig:GEF5}\includegraphics[width=60mm]
   {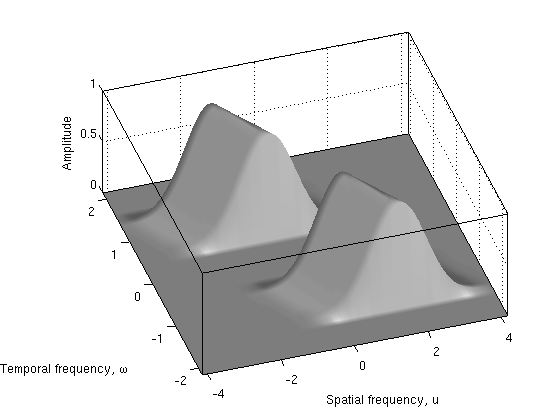}} 
  \subfloat[``\textit{(d) minus (b)}'']{\label{fig:GEF6}\includegraphics[width=60mm]{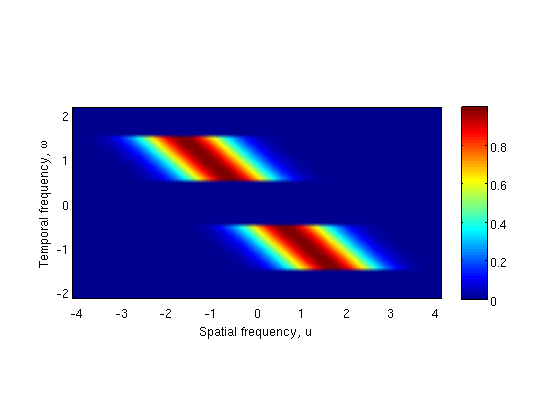}}
 \label{figur_GEF}\caption{Construction of temporal frequency bandpass Gabor-Einstein wavelet. Each surface plot in the right column is of the same Gabor-Einstein element as its colormap neighbor on the left. Parameters used: $\sigma=1,~u_0=1.15$.}

\label{fig:GE_band}
 \end{figure}

\subsection{Temporal Frequency Filter Property Distribution}
\label{subsec:temp_fil_sim}

\begin{figure}[htp]
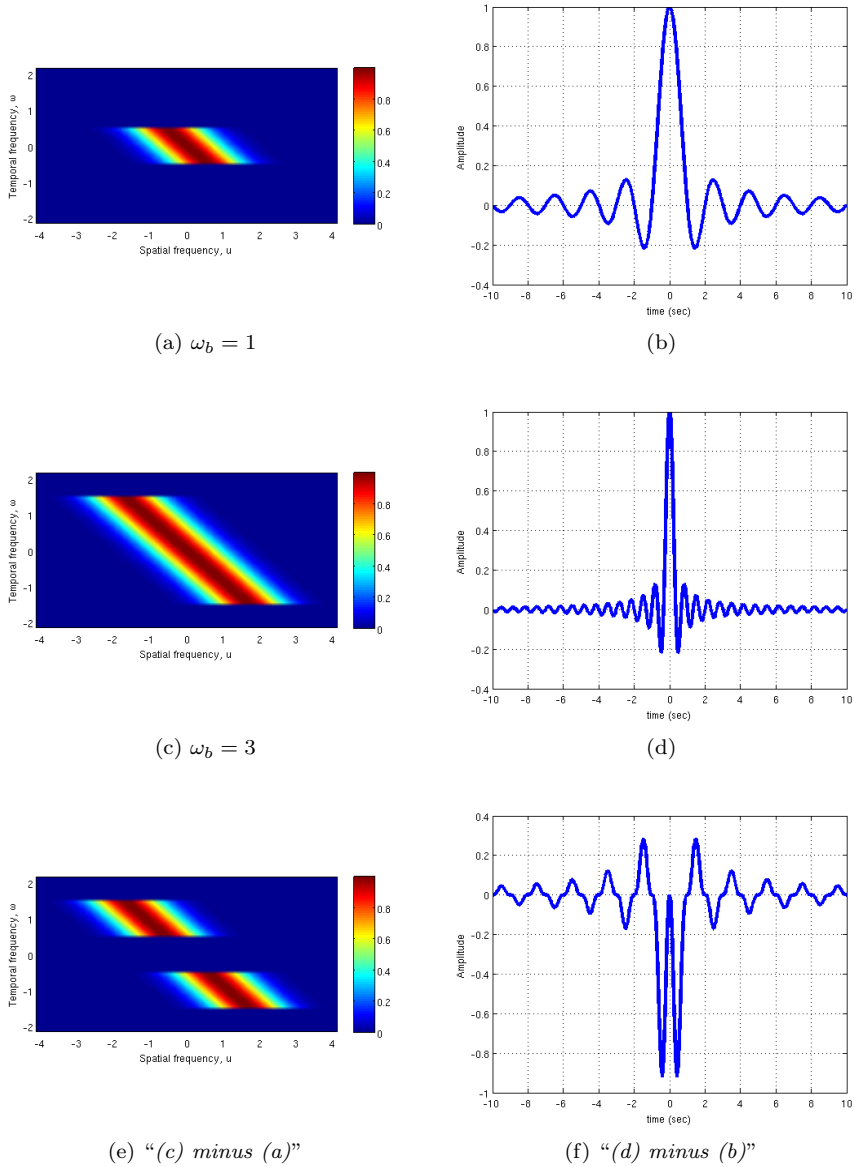


  \centering
  \subfloat[$\omega_b=1$]{\label{fig:STS1}\includegraphics[width=60mm]
   {GEF_band_pcolor1.png}} 
  \subfloat[]{\label{fig:STS2}\includegraphics[width=60mm]
{sinc_1D.png}}
\\
  \subfloat[$\omega_b=3$]{\label{fig:STS3}\includegraphics[width=60mm]
   {GEF_band_pcolor2.png}} 
  \subfloat[]{\label{fig:STS4}\includegraphics[width=60mm]
{sinc_1D_w3.png}}
\\
  \subfloat[``\textit{(c) minus (a)}'']{\label{fig:STS5}\includegraphics[width=60mm]
   {GEF_band_pcolor3.png}} 
  \subfloat[``\textit{(d) minus (b)}'']{\label{fig:STS6}\includegraphics[width=60mm]{sinc_1D_w3-w1}}
 \label{figur_STS}\caption{Progressive complexity: The bandpass temporal frequency filters such as (e) are necessarily superpositions of two or more Gabor-Einstein wavelets. The profiles on the right are the temporal response weights probed at the spatial origin. Note that the pattern is more complex in the bandpass temporal frequency filter. Parameters used: $\sigma=1,~u_0=1.15$.}

\label{fig:GE_striatotemporal spectrum}
 \end{figure}

Figure~(\ref{fig:sinc and fourier}) shows the essential property which the Gabor-Einstein wavelet inherits from the sinc function. The sinc function's fourier transform is a lowpass filter. Higher frequency sinc functions have wider bandwidth. Bandpass filters are formed by taking the difference between sinc functions of different frequency as illustrated in Figure~(\ref{fig:sinc and fourier}). The Gabor-Einstein wavelet's fourier transform has a step function factor inherited from the sinc function. This allows it explain the particular distribution of lowpass to bandpass temporal frequency filter properties of V1 to MT neurons (Foster et al 1985; DeAngelis et al 1993b; Hawken et al 1996) in a manner innately representative of the motion-processing stream's neuronal hierarchy. For simplicity of illustration, Figure~(\ref{fig:GE_Fourier}) shows a 2D (x,t) Gabor-Einstein element. Its fourier transform is shown in Figs (\ref{fig:GE_Fourier}a) and (\ref{fig:GE_Fourier}b), while its (x,t) domain representation is shown in Fig (\ref{fig:GE_Fourier}c). The lowpass temporal frequency nature of the fourier representation is apparent, as the support straddles the zero. Figure~(\ref{fig:GE_band}) shows the construction of temporal frequency bandpass Gabor-Einstein wavelets.  As illustrated, the temporal frequency bandpass Gabor-Einstein element results from the difference of two temporal frequency lowpass Gabor-Einstein elements. Fig~(\ref{fig:GE_band}a) and Fig~(\ref{fig:GE_band}b) are plots of Equation~(\ref{Gabor-Einstein-Fourier}), i.e. they are the fourier transform of the Gabor-Einstein wavelet described by Equation~(\ref{Gabor-Einstein}). We label this basis element ``$\omega_b=1$'', meaning it is the zero-centered lowpass temporal frequency filter whose temporal frequency bandwidth is one. Accordingly, Fig~(\ref{fig:GE_band}c) and Fig~(\ref{fig:GE_band}d) are plots of the ``$\omega_b=3$'' Gabor-Einstein basis element, i.e. they are plots of the zero-centered lowpass temporal frequency filter whose temporal frequency bandwidth is three. Fig~(\ref{fig:GE_band}e) and Fig~(\ref{fig:GE_band}f) plot the bandpass temporal frequency filter basis element obtained by taking the difference of the ``$\omega_b=1$'' and ``$\omega_b=3$'' basis elements. The left-hand column of Figure~(\ref{fig:GE_striatotemporal spectrum}) plots the Gabor-Einstein basis elements of Fig~(\ref{fig:GE_band}), while the right hand column plots the corresponding temporal response probed at the spatial origin of the receptive field. There is an increasing complexity of the temporal waveform structure as one progresses from lowpass to bandpass basis element. 

\begin{figure}[htp]

  \centering
  \subfloat[]{\label{fig:GEF_sum1}\includegraphics[width=60mm]
   {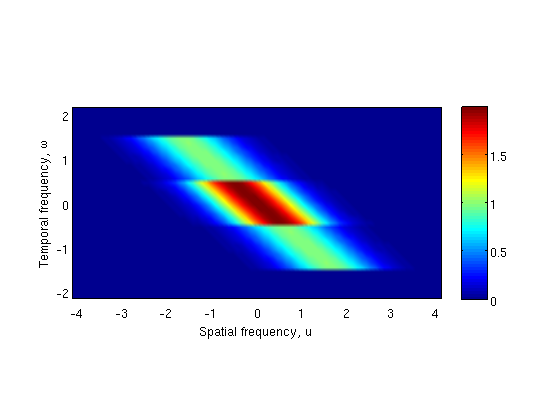}} 
  \subfloat[]{\label{fig:GEF_sum_p}\includegraphics[width=60mm]
   {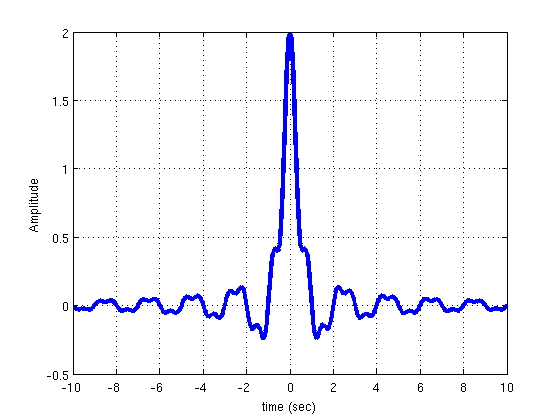}}\\
  \subfloat[]{\label{fig:GEF_sum_s}\includegraphics[width=60mm]
   {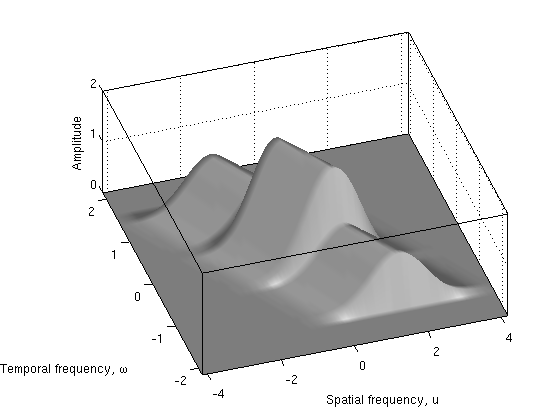}}

\caption{A superposition of two lowpass filters which yields a hierarchically more complex neuron which is nonetheless low pass. Specifically, the above is a plot of ``$\omega_b=1$'' + ``$\omega_b=3$''. Parameters used: $\sigma=1,~u_0=1.15$.}
\label{fig:GE_sum}
 \end{figure}

All pure Gabor-Einstein basis elements are lowpass temporal frequency filters. On the other hand, the bandpass temporal frequency filter property necessarily results from summation of Gabor-Einstein basis elements. Hence bandpass temporal elements are necessarily complex (not pure). However, not all lowpass temporal frequency filters are pure; and not all basis summations yield bandpass temporal frequency filters. For instance, Figure~(\ref{fig:GE_sum}) shows the sum of two pure Gabor-Einstein basis elements which yield another lowpass temporal frequency filter. The temporal waveform of the composite element is indeed more complex than that of its two pure constituents, however, it appears less complex than the composite temporal waveform of Fig~(\ref{fig:GE_band}).

\section{Discussion}
\label{sec:discussion}

The neurons in the V1 to MT dorsal stream are highly specialized towards motion detection. Furthermore, they are embedded in a hierarchical order of increasing motion specialization. The earliest (least specialized) members of the chain are the non-speed tuned V1 cells, while the most specialized members are the MT cells. Naturally, LGN cells precede input layer V1 cells both anatomically as well as in level of motion specialization. Similarly, cells located in the medial superior temporal (MST) area and the parietal lobe are further along than MT cells both anatomically and in motion specialization. We have focused our current study on the V1 to MT spectrum.

We set out to obtain a good mathematical representation for the receptive fields of these neurons. The term \textit{good} here implies that the representation should faithfully capture the salient features of the motion specialization spectrum and serve as an effective label or address schema for neurons in the motion processing stream. For instance, it is known that MT neurons receive input from V1 neurons and most directly from V1-to-MT projectors, therefore in a good basis, the representation for an MT neuron should result from some combination of representations of some V1 neuron(s). Furthermore, the resulting MT representation should naturally manifest known attributes of MT neurons.

Appropriate comparison and discussion of the current work and related work requires a more precise language than currently available. Hence, we begin this discussion by introducing a simple framework for classifying and naming the components and hierarchical levels of receptive field models.

\subsection{Receptive Field Model Classification}
\label{subsec:model_framework}

In the literature, there is currently no clear categorization of the various mathematical descriptions of receptive fields. Ironically, the laudable increase in discovery of receptive field attributes and modulation mechanisms may have led to further conceptual entanglement of the mathematical forms of receptive fields. Here, we propose a simple framework to organize and classify receptive field descriptions and briefly review some basic concepts. This should help place the current work, the Gabor-Einstein model, in the appropriate context; and furthermore, should help organize and direct receptive field research. We will identify a receptive field by its constituents and method of construction. A receptive field can be constructed using two essential things:

\begin{enumerate}
 \item Primary input receptive fields
\item Input combination scheme
\end{enumerate}

The primary input receptive field is the response function of the neuron at the lowest level on the hierarchy of study. For instance, when studying the motion-processing stream between V1 and MST, the primary input receptive field is the response function of a V1 input layer neuron. An input combination scheme is a model for the network of neurons projecting onto the study neuron. For instance, a common scheme is linear summation of weighted V1 responses to yield an MT neuron. The input combination scheme can be classified according to the stage at which input processing, if any, occurs. Three classes arise:

\begin{enumerate}
 \item Pre-processing
\item Intra-processing
\item Post-processing
\end{enumerate}

Pre-processing is modification of the input responses prior to combining them. For instance, V1 elements are squared in the motion-energy model, after which they are summed to yield the MT response~\citep{adbe1985,adbe1986}. Intra-processing is modification of the input responses during the combination process. For example, the Soft MAX scheme~\citep*{lafe2004,ripo1999,nose1995}. Post-processing is modification of the response after input combination. For instance, divisive normalization schemes~\citep*{he1992,he1993,sihe1998,tshu2010}. The divisive normalization scheme, proposed by Heeger in 1992, is a neuron population based model of V1 to MT function. In that model, each neuron's behavior is modulated by that of a population module of which it was a member. Specifically, its manifest response is its native response divided by the net sum of the responses of the other neurons in its population module plus a regularization factor. The model bears resemblance to receptor models in pharmacokinetics, where the regularization factor plays the role of a half-saturation constant. The third categorization attribute is the hierarchical level from which input originates relative to the study neuron. The three following classes arise:

\begin{enumerate}
 \item Feed-forward
\item Isostratal
\item Feedback
\end{enumerate}

Feed-forward implies the input comes in from a lower hierarchical level. For instance V1 projecting to MT. Isostratal implies that the input comes in from the same level. For instance, normalization of MT response by neighborhood neuron pool~\citep*{he1992,he1993,sihe1998,tshu2010}. Feedback implies the input comes in from a higher level in the hierarchy. Of note, most receptive fields likely receive contributions from all three types of inputs. In light of the above framework, the Gabor-Einstein model is a primary input receptive field for the V1 to MT motion processing stream. It is compatible with any of the aforementioned input combination schemes, hierarchical input routing schemes, and input processing schemes. The work presented in this paper focused mainly on the primary input receptive field, and for simplicity of illustration, assumed a basic (i.e. linear) input combination scheme. Certain important phenomena cannot be explained at the level of early (e.g. input layer V1) primary input receptive fields alone. Next, we discuss some aspects of such phenomena including end-stopping, a short selection of non-linearity mechanisms, and contrast-modulated speed tuning.

\subsection{Related Work}
\label{subsec:related_work}
A key distinguishing feature of the Gabor-Einstein model is how it naturally represents the distribution of lowpass to bandpass temporal frequency filtering properties along the V1 to MT motion specialization hierarchy. Ratio models are a class of models with a different objective. The ratio models aim to mechanistically model the way speed tuned units arise from combinations of non-speed tuned bandpass and low pass temporal frequency filters. These models are based on the idea that stimulus temporal frequency is proportional to the ratio of outputs of spatiotemporally separable bandpass to lowpass neuronal outputs~\citep*{tosh1973,adbe1986,peth2002,pe2004,pe2005}.

Various summation-based input combination schemes along the V1 to MT stream have been used. Heeger and colleagues proposed a two stage weighted linear summation model in which MT fields resulted from weighted sums of V1 fields~\citep{he1992,he1993,sihe1998}. Sereno as well as Nowlan \& Sejnowski proposed multilevel neural network reinforcement learning models in which MT neuron response was a function of a linear summation of V1 responses~\citep{se1993,nose1995}. Tsui et al proposed a model in which MT response was obtained by a soft-max weighted summation of V1 inputs~\citep*{tshu2010}. Each of these schemes differs fundamentally in approach from the Gabor-Einstein model. Namely, we focus on the receptive field of a single neuron along the V1 to MT pathway, implicitly representing the network spatiotemporal structure in the inherent attributes of the basis itself. Specifically, the natural representation of temporal frequency filter distribution along the V1 to MT spectrum. In the Gabor-Einstein basis, the basis itself mandates summation as a means of ascending up the specialization hierarchy. This is a gratifying result, and supports the notion that the sinc function is a natural way to describe the V1 to MT spectrum.

\subsection{Higher-order Phenomena}
\label{subsec:higher_order}

Here, we discuss certain higher order phenomena whose description exceeds the scope of single early neuron receptive fields.

\subsubsection{End-stopping}
End-stopping is a center-surround type phenomenon in which ---for instance--- a full length bar stimulus results in a submaximal response, while some shorter bar stimulus generates maximal response. In other words, the excitatory portion of the receptive field is shorter than the full length of the receptive field. Typically, the periphery of the receptive field, i.e. its \textit{end}, elicits suppression (or ``stopping''). Hence neurons exhibiting such behavior are said to be ``end-stopped''.

Jones and colleagues found that majority (94\%) of V1 cells in the macaque were end-stopped to various degrees~\citep{jogr2001}. Moreover, the most prominently end-stopped V1 cells have been found in layer 4B which is known to be the dominant projection layer to MT~\citep{scha2001}. Furthermore, Tsui et al have argued a role for V1 end-stopping in MT motion integration and solution to the aperture problem~\citep*{tshu2010}. It seems therefore that a complete model of V1 to MT neurons must account for end-stopping. End-stopping phenomena would be difficult to explain at the single input layer V1 neuron stage, since it likely involves isostratal interactions. Tsui and colleagues modeled end-stopping by specifying the normalization pool as surround cells arranged along the orientation axis of a center cell. With this spatially-specific implementation of normalization, surround cell activity yielded end-stopping~\citep{tshu2010}. The response was then fit to a saturation kinetics model equivalent to Heeger's divisive normalization model~\citep{he1992}. Regarding center-surround phenomena such as end-stopping, the Gabor-Einstein wavelet can be considered a \textit{classical receptive field}. A scheme similar to that of Tsui et al can be used to implement end stopping using the Gabor-Einstein wavelet as both center and surround, or in hybrid with other wavelet transform as either center or surround. Of note, other schemes have also been used to implement end-stopping. For instance, Skottun proposed an orientation-modulated cancellation strategy~\citep{sk1998}. The Gabor-Einstein wavelet is compatible with this scheme as well.

\subsubsection{Nonlinearity Mechanisms}

The perception of motion is likely a higher order phenomenon that emerges out of the elaborate nonlinear connection network of which MT is only a part. Some of these non-V1 areas from which MT receives input include LGN ~\citep*{sipa2004,naly2006}, superior colliculus~\citep*{rogr1990,bewu2010,bewu2011}, and extrastriate regions~\citep*{mava1983,boun1990,unde1986}. Some combination of these non-V1 inputs may explain the persistence of visual responsiveness and direction selectivity after V1 lesions or cooling ~\citep*{gisa1992,azpa2003,roal1989,rogr1990}. Indeed several researchers have drawn specific attention to the necessity of non-linear models. Moreover, these nonlinearities are not just a feature of late (specialized) visual neurons, but have been demonstrated very early in the visual pathway. For example, Schwartz et al pointed out the need for nonlinear retinal ganglion cell receptive field models~\citep{scok2012}. And Rosenberg et al showed that LGN Y cells utilize a nonlinear mechanism to represent interference patterns~\citep*{rohu2010}. Fine mechanistic modeling of the single V1 to MT neuron's receptive field's most salient features is certain to yield valuable physiological insight into the myriad non-V1 connections to MT.

\subsubsection{Contrast-Modulated Effects}
Contrast-modulated speed tuning is likely unexplainable at the level of early primary input receptive fields of single neurons. Existing explanations of such phenomena appear to necessarily involve nonlinear isostratal interactions. For instance, divisive normalization~\citep*{he1992,he1993,sihe1998}. 

 In general, the effects of contrast on the receptive field of V1 to MT neurons can safely be assumed to be highly modulated by neighboring neurons. Significant data exists to support such population encoding. For instance, DeAngelis et al showed that end and side stopping were strongest along the preferred excitatory orientation, yet superimposed suppressive stimuli was much more broadly tuned than the classical receptive field, suggesting a modulatory pool of neurons~\citep{defr1994}. Such neuron pools also likely mediate the effects of contrast on speed tuning. Krekelberg et al found that most MT cells in alert macaques preferred lower speeds for lower stimulus contrast~\citep*{krva2006}. Priebe et al found that a significant population of V1 cells have contrast-modulated speed tuning curves~\citep*{prli2006}. Livingstone and Conway found that the speed tuning curves of most V1 neurons in alert macaques were dependent on contrast stimuli. Specifically, for lower contrast stimuli, V1 cells shifted their tuning curves to lower speeds; while for higher contrast stimuli, the opposite was observed~\citep*{lico2007}. Psychophysical studies also show that both high contrast stimuli and high spatial frequency stimuli are associated with higher perceived speeds~\citep*{brmo2011}. The increase in preferred speed with contrast is a phenomenon that likely involves nonlinear isostratal interactions. As such, early single neuron receptive field models in isolation are unable to explain such phenomena. Nonetheless the primary input receptive field is the theoretical and physiological foundation of higher order phenomena. Hence it is essential to have a physiologically sound model such as the Gabor-Einstein wavelet which naturally generates the hierarchical specialization structure of the motion stream. The Gabor-Einstein wavelet can be readily plugged-in to divisive normalization schemes which may be necessary to explain contrast-modulated phenomena.

\begin{figure}[ht]
\begin{center}
\scalebox{.49}
{\includegraphics{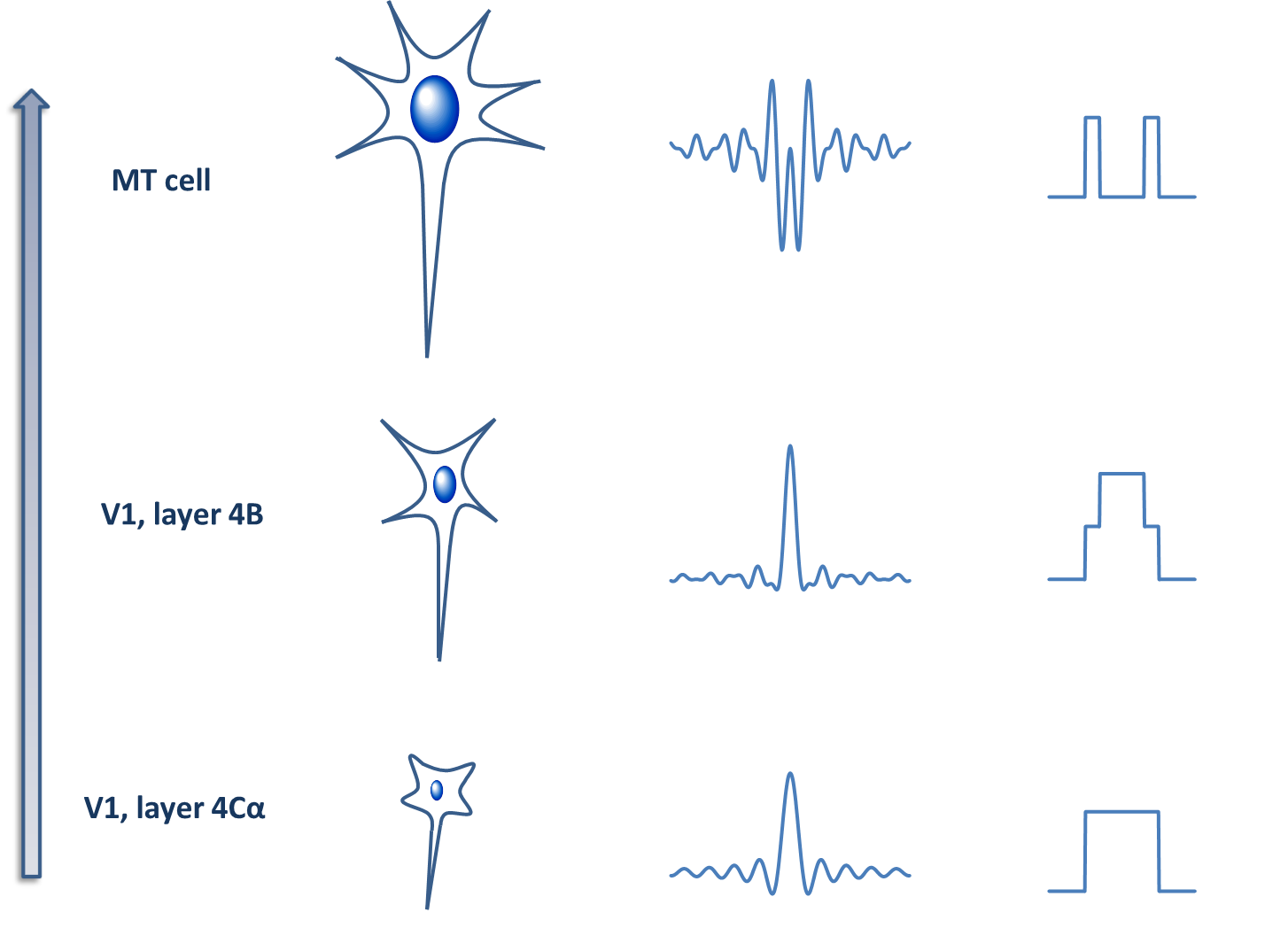}}
\end{center}
\caption{A schematic illustration of the V1 to MT motion specialization spectrum. The arrow gradient represents increasing motion specialization along the spectrum. In the second column, the neurons depict progressively complex specializations in size and arborization patterns along the spectrum. The third column depicts the temporal profile probed at the spatial origin of the receptive field. The waveform progresses from a pure sinc function in the upstream layers to composite forms downstream. The rightmost column depicts the increased proportion of bandpass to lowpass temporal frequency filters as one ascends the hierarchy.}
\label{fig:V1-MT}
\end{figure}

\section{Conclusion}
\label{sec:conclusion}

In this paper, we introduced the Gabor-Einstein Wavelet, a new family of functions for modeling the receptive fields of neurons in the V1 to MT motion processing stream. We showed that the way space and time are mixed in the visual cortex has analogies to the way they are mixed in the special theory of relativity. We therefore constrained the Gabor-Einstein model to a relativistically invariant wave carrier and to the minimum possible number of parameters. These constraints yielded a sinc function wave carrier with energy-momentum relation as argument. The model innately and efficiently represents the temporal frequency filtering property distribution along the motion processing stream. Specifically, on the V1 end of the stream, the neuron population has an equal proportion of lowpass to bandpass temporal frequency filters; whereas on the MT end, they have mostly bandpass temporal frequency filters. From our analysis and simulations, we showed that the distribution of temporal frequency filtering properties along the motion processing stream is a direct effect of the way the brain jointly encodes space and time. We uncovered this fundamental link by demonstrating an analogous mathematical structure between the special theory of relativity and the joint encoding of space and time in the visual cortex.  The Gabor-Einstein model and the experiments it motivates will provide new physiological insights into how the brain represents visual information. 

\newpage
\section*{Acknowledgements}
The author thanks Rudi Weikard, Marius Nkashama, John Mayer, Ian Knowles, Xiaobai Sun, and Peter Blair for helpful suggestions on how to approach the fourier transform of the sinc function of multidimensional argument. He thanks Greg Schwartz and Ari Rosenberg for helpful discussion on nonlinearities in visual receptive fields. He thanks his mentors and role models in Ophthalmology at Howard University and the Washington area for their dedication to training residents: Robert A. Copeland Jr., Leslie S. Jones, Bilal Khan, Earl Kidwell, Janine Smith-Marshall, David Katz, William Deegan III, Melissa Kern, Ali Ramadan, Frank Spellman, Reggie Sanders, and Michael Rivers, Emily Chew, Brian Brooks, and Wai Wong. He thanks Claude L. Cowan Jr whose excellence inspired him to pursue a career in Medical Retina. He thanks his co-residents at Howard Ophthalmology for the wonderful experience which resulted from their enthusiasm for patient care and collegial learning: Sir Gawain Dyer, Salman J. Yousuf, Animesh Petkar, Ninita Brown, Mona Kaleem, Mikelson MomPremier, Saima Qureshi, Nikisha Richards, Neal Desai, Chris Burris, Natasha Pinto, Usiwoma Abugo, Chinwe Okeagu, and Katrina del Fierro. He thanks Richard Mooney, Michael Platt, Pate Skene, Fan Wang, Vic Nadler, and Kafui Dzirasa for supporting his membership to the Society for Neuroscience. He thanks Susan Elner, Mark W. Johnson, John R. Heckenlively, Paul P. Lee and all the amazing faculty at the University of Michigan--Ann Arbor (W.K. Kellogg Eye Center) for awarding me the clinical fellowship training position in Medical Retina. He thanks Stuart Fine and the board of directors of the Heed ophthalmic foundation (Nicholas J. Volpe, Stephen McLeod, Joan Miller, Eduardo Alfonso, David Wilson, Julia Heller, and Froncie Gutman) for inviting him to participate in the 8th Annual Heed resident's retreat in Chicago Illinois. It was a delight to be at that retreat. He thanks Eydie Miller-Ellis, Mildred Olivier, and the the Rabb Venable Excellence in Research Program Board for selecting him as a participant and for encouraging a career in academia.

\newpage
\section*{Author Biography}

{\footnotesize{
Dr. Stephen G. Odaibo is Chief Scientist and Founder of Quantum Lucid Research Laboratories. He received the 2005 Barrie Hurwitz Award for Excellence in Clinical Neurology from Duke University School of Medicine where he topped the class in Neurology. Dr. Odaibo is a Mathematician, Computer Scientist, Physicist, and Physician. He obtained a B.S. in Mathematics (UAB, 2001), M.S. in Mathematics (UAB, 2002), M.S. in Computer Science (Duke, 2009), and Doctor of Medicine (Duke, 2010). Dr. Odaibo completed his internship in Internal Medicine at Duke University Hospital and is currently an Ophthalmology resident at Howard University Hospital in Washington DC. He is author of the book, ``Quantum Mechanics and the MRI Machine'' (Symmetry Seed Books, Oct 2012). He invented the Trajectron method with which he provided the first quantitative demonstration of non-paraxial light bending within the human cornea. His other awards and recognitions include: He won the 2013 Best Resident Research Presentation Award at the 23rd Annual Washington Retina Symposium; in 2012 he was selected as a Featured Alumnus of the Mathematics Department at UAB; and his cornea paper was selected by MIT Technology Review as one of the best papers from Physics or Computer science submitted to the arXiv the first week of Oct 2011. Dr. Odaibo's research interests are at the fusion of Mathematics, Computer Science, Physics, and the Biomedical Sciences, with a special focus on the representation of motion in the mammalian visual cortex. His clinical interests are in the diagnoses and medical management of retinal disease. In the 2014-2015 academic year, Dr. Odaibo will be a Medical Retina fellow at the University of Michigan--Ann Arbor. In his spare time he enjoys time with his wife, family and friends, and studying the bible.}}

\end{document}